\documentclass[draft]{agujournal2019}
\usepackage{url} %this package should fix any errors with URLs in refs.
\usepackage{lineno}
\usepackage[inline]{trackchanges} %for better track changes. finalnew option will compile document with changes incorporated.
\usepackage{soul}
\usepackage{amsmath,amssymb}
\usepackage{xspace}
\usepackage{subfig}
\usepackage{comment}
\usepackage{multirow}
\usepackage{booktabs}
\usepackage{subfig}
\usepackage{caption}

\newcommand{\dtec}{$\delta\rm{TEC}$\xspace}

\draftfalse

\journalname{Geophysical Research Letters}

\begin{document}

%% ------------------------------------------------------------------------ %%
%  Title
%
% (A title should be specific, informative, and brief. Use
% abbreviations only if they are defined in the abstract. Titles that
% start with general keywords then specific terms are optimized in
% searches)
%
%% ------------------------------------------------------------------------ %%

% Example: \title{This is a test title}

\title{Spectral Analysis of ionospheric density variations measured with the large radio telescope in the Low-Latitude region}
%% ------------------------------------------------------------------------ %%
%
%  AUTHORS AND AFFILIATIONS
%
%% ------------------------------------------------------------------------ %%

% Authors are individuals who have significantly contributed to the
% research and preparation of the article. Group authors are allowed, if
% each author in the group is separately identified in an appendix.)

% List authors by first name or initial followed by last name and
% separated by commas. Use \affil{} to number affiliations, and
% \thanks{} for author notes.
% Additional author notes should be indicated with \thanks{} (for
% example, for current addresses).

% Example: \authors{A. B. Author\affil{1}\thanks{Current address, Antartica}, B. C. Author\affil{2,3}, and D. E.
% Author\affil{3,4}\thanks{Also funded by Monsanto.}}

\authors{Sarvesh Mangla\affil{}, Abhirup Datta\affil{}}

% \affiliation{1}{First Affiliation}
% \affiliation{2}{Second Affiliation}
% \affiliation{3}{Third Affiliation}
% \affiliation{4}{Fourth Affiliation}

\affiliation{}{Department of Astronomy, Astrophysics and Space Engineering, Indian Institute of Technology Indore, Madhya Pradesh, 453552, India}
%(repeat as many times as is necessary)

%% Corresponding Author:
% Corresponding author mailing address and e-mail address:

% (include name and email addresses of the corresponding author.  More
% than one corresponding author is allowed in this LaTeX file and for
% publication; but only one corresponding author is allowed in our
% editorial system.)

% Example: \correspondingauthor{First and Last Name}{email@address.edu}

\correspondingauthor{Sarvesh Mangla}{mangla.sarvesh@gmail.com}

%% Keypoints, final entry on title page.

%  List up to three key points (at least one is required)
%  Key Points summarize the main points and conclusions of the article
%  Each must be 140 characters or fewer with no special characters or punctuation and must be complete sentences

% Example:
% \begin{keypoints}
% \item	List up to three key points (at least one is required)
% \item	Key Points summarize the main points and conclusions of the article
% \item	Each must be 140 characters or fewer with no special characters or punctuation and must be complete sentences
% \end{keypoints}

\begin{keypoints}
  \item GMRT can demonstrate an order of magnitude better sensitivity than GNSS-based TEC measurements in characterizing ionospheric fluctuations.
  \item The spectral analysis technique used with GMRT can detect multiple MSTIDs and smaller-scale structures simultaneously.
  \item GMRT can detect ionospheric variations as small as 10 km. The study also showed waves changing direction unexpectedly during sunrise time.
\end{keypoints}

%% ------------------------------------------------------------------------ %%
%
%  ABSTRACT and PLAIN LANGUAGE SUMMARY
%
% A good Abstract will begin with a short description of the problem
% being addressed, briefly describe the new data or analyses, then
% briefly states the main conclusion(s) and how they are supported and
% uncertainties.

% The Plain Language Summary should be written for a broad audience,
% including journalists and the science-interested public, that will not have 
% a background in your field.
%
% A Plain Language Summary is required in GRL, JGR: Planets, JGR: Biogeosciences,
% JGR: Oceans, G-Cubed, Reviews of Geophysics, and JAMES.
% see http://sharingscience.agu.org/creating-plain-language-summary/)
%
%% ------------------------------------------------------------------------ %%

%% \begin{abstract} starts the second page

\begin{abstract}
    The low-latitude ionosphere is a dynamic region with a wide range of disturbances in temporal and spatial scales. The Giant Metrewave Radio Telescope (GMRT) situated in the low-latitude region has demonstrated its ability to detect various ionospheric phenomena. It can detect total electron content (TEC) variation with precision of 1\,mTECU and also can measure TEC gradient with an accuracy of about $\rm7\times 10^{-4}\,TECU\,km^{-1}$. This paper describes the spectral analysis of previously calculated TEC gradient measurements and validates them by comparing their properties using two bands. The analysis tracked individual waves associated with medium-scale traveling ionospheric disturbances (MSTIDs) and smaller waves down to wavelengths of $\sim$\,10\,km. The ionosphere is found to have unanticipated changes during sunrise hours, with waves changed propagation direction as the sun approached the zenith. Equatorial spread\,$F$ disturbances during sunrise hours is observed, along with smaller structures moving in the same direction. 
\end{abstract}

\section*{Plain Language Summary}

The Earth's ionosphere can limit exploring sub-GHz frequencies of the sky and introduces an extra phase term that is difficult to calibrate. The same calibration data can be used to study the Earth's ionosphere more precisely than conventional probes. Radio interferometry is a technique for studying astronomical sources and Earth's ionosphere by measuring the spatial coherence function of multiple elements. The GMRT is a unique instrument for exploring the equatorial ionosphere region. This study used dual-band observations of a bright radio source with the GMRT to explore the Equatorial Ionization Anomaly region. The GMRT can detect variations in total electron content and measure TEC gradient with high accuracy. Spectral analysis was performed on TEC gradient measurements to track individual waves associated with medium scales traveling ionospheric disturbances and smaller waves up to wavelengths of about $\sim$\,10\,km. 
The results showed unexpected changes in the ionosphere during sunrise hours and observed large plasma irregularities and smaller structures moving in the same direction.

%% ------------------------------------------------------------------------ %%
%
%  TEXT
%
%% ------------------------------------------------------------------------ %%

%%% Suggested section heads:
% \section{Introduction}
%
% The main text should start with an introduction. Except for short
% manuscripts (such as comments and replies), the text should be divided
% into sections, each with its own heading.

% Headings should be sentence fragments and do not begin with a
% lowercase letter or number. Examples of good headings are:

% \section{Materials and Methods}
% Here is text on Materials and Methods.
%
% \subsection{A descriptive heading about methods}
% More about Methods.
%
% \section{Data} (Or section title might be a descriptive heading about data)
%
% \section{Results} (Or section title might be a descriptive heading about the
% results)
%
% \section{Conclusions}

\section{Introduction}

Radio-frequency arrays, especially those that operate in the low-frequency range ($\lesssim 1$ GHz), are a potent but comparatively underutilised tool for remote sensing. They are mostly used to observe cosmic sources and were built as synthesis telescopes. To mitigates the effects of the ionosphere, radio interferometers need detailed calibration schemes. However, the same calibration data is seldom used to detect and study the Earth's ionosphere.

Radio interferometers measure `Spatial Coherence Function' \cite{thompson_book}, where an additional phase term is introduced because of the ionosphere. This phase is proportional to the difference in TEC along the line of sight between two-array elements. Thus, these extra phase term can be easily converted to differential TEC (\dtec) with an accuracy of $10^{-3}$ TECU \cite{Mangla2022MNRAS.513..964M, mev16} or better \cite{Helm2012RaSc...47.0K02H_temporal}. Interferometers are effective tools for studying ionospheric dynamics due to their increased sensitivity to measure TEC gradients compared to traditional probes such as Radars, ionosondes, and the Global Navigation Satellite System (GNSS). Recently, they are also used to study Travelling Ionosphere Disturbances (TIDs; \citeA{Jac1992A&A...257..401J}) and even ionospheric scintillations \cite{Rich2020JSWSC..10...10F}. \par
Using the Very Large Array (VLA), \citeA[ and references within]{Jac1992A&A...257..401J} measured phases from individual antennas by observing a single-bright source and further performing spectral analysis to detect TIDs primarily associated with gravity waves. Additionally, \citeA{Helm2012RaSc...47.0L02H_spectral} utilized a similar spectral analysis on TEC gradient observations with the VLA to detect several small-scale structures and medium-scale TIDs (MSTIDs). It was observed that the smaller scale disturbances propagated in the same direction as the MSTIDs. Later, using VLA Low-frequency Sky Survey, \citeA{Helmboldt2012RaSc...47.5008H} detected many wavelike disturbances at 74\,MHz, by measuring the positional shift of several bright radio galaxies to obtain fluctuation spectra for the TEC gradient. Studying these spectra gave a detailed account from ionospheric intraday variation to seasonal variation where turbulent activity appear to be significant during the summer nighttime, winter daytime and during sunset in the spring near mid-latitude region. \par

By utilising the night-time observation of the wide fields of view of the Murchison Widefield Array (MWA), \citeA{Loi_2015} conducted a power spectrum analysis of ionospheric fluctuations (measured by positional offsets of several radio sources) to reveals field-aligned irregularities (large electron irregularities elongated along the geomagnetic field lines; \citeA{Loi2015GeoRL..42.3707L, Loi2016RaSc...51..659L}) and other wave-like phenomena including TIDs \cite{Loi2016JGRA..121.1569L}. Later, \citeA{Helm2020RaSc...5507106H} utilized the GLEAM Survey \cite{Wayth2015PASA...32...25W} of the MWA, to generate images of ionospheric structures present during the observations. Furthermore, spectral analysis of these images provided evidence of distinct features of ionospheric activity, including the generation of MSTIDs in conjunction with sporadic E layer (Es) events during nighttime. A recent analyses with the LOw Frequency ARray (LOFAR), GNSS and ionosondes, \citeA{Rich2020JSWSC..10...10F} revealed breaking down of large-scale ionospheric structure into smaller-scales, giving rise to the scintillation. \par

Although, above mentioned studies are done with radio telescopes, these studies are limited to mid-latitude region due to geographical location constraints. This makes the Giant Meterwave Radio Telescope (GMRT; geophysical latitude and longitude: $19^{\circ} \, 05' $ N and  $74^{\circ} \, 03'$ E; geomagnetic latitude: $10^{\circ} \, 40'$ N) a unique instrument to study the ionosphere in low-latitude region. It is important to note that, GMRT is located in the geophysically sensitive region between the magnetic equator and the northern crest of the Equatorial Ionization Anomaly (EIA; \citeA{Appl1946Natur.157..691A}) in the Indian longitude sector. \par
In our most recent study \cite{Mangla2022MNRAS.513..964M}, we explored the techniques for deriving ionospheric data from calibration of the GMRT data. We successfully achieved a typical \dtec precision of $10^{-3}$ TECU while monitoring a bright radio source. We also demonstrated methods for measuring the TEC gradient with an accuracy of about $\rm7\times 10^{-4}\,TECU\,km^{-1}$ since arrays like this are essentially only sensitive to the TEC gradient. Here, we propose spectral analysis methods for these TEC gradient observations in an effort to further this project. We will show that these techniques can find and describe a number of medium- to small-scale phenomena, as well as give a general statistical description of the spectrum of TEC variations. \par
The present paper is outlined as follows: in  Sec. \ref{sec:observation} we summarise our observations with GMRT and computed TEC gradient measurements. In Sec. \ref{sec:spectalanalysis}, we outlined the spectral analysis method to estimate the properties of ionospheric structures using GMRT observations and provide limitation of how much smaller waves may be detected using this method. Finally, in  Sec. \ref{sec:conclusion}, we conclude our results and discussion. 

\section{Observation and TEC gradient measurements}
\label{sec:observation}

A nearly 9-hour observation with GMRT of a cosmic radio source (3C\,68.2) served as the basis for this analysis. The observations were taken simultaneously at 235 and 610\,MHz on the night of August 5 and 6, 2012. The observations consisted of each `scan' (block of time) of nearly one hour, except scan number four, which is nearly five hours continuous observation during mid-night to post-sunrise hours with a temporal sampling of 0.5\,s. Furthermore, a moderate level of geomagnetic activity ($K_p$ index $\sim$ 1-3) and a moderate level of solar activity (F10.7 = 137.9 SFU; 1SFU = $\rm 10^{-22}\,W\,m^{2}\,Hz^{-1}$) were both present throughout the observation. \par

Measured TEC gradients using the GMRT are extensively described in detail in \citeA{Mangla2022MNRAS.513..964M}. In brief summary, the ionosphere introduces an extra phase term, proportional to the difference in TEC (or \dtec), measured along the line of sight of a pair of antennas. This makes the radio interferometers sensitive to TEC gradients, rather than the absolute TEC value. However, due to the Y-shape of the GMRT (consisting of northeastern, northwestern, and southern arms), it is challenging to accurately measure the TEC gradient and perform Fourier inversion on it. \citeA{Mangla2022MNRAS.513..964M} utilised two methods (previously discussed by \citeA{Helm2012RaSc...47.0K02H_temporal}) to address this issue. The first method involves estimating the full 2-D TEC gradient over the array, utilizing a second- and third-order polynomial (Taylor series) at each time step. This method can recover the large, long-period disturbances that are present during the observation. \par
The second approach involves computing the projection of the TEC gradient along each arm of the GMRT at each time step. This method provides limited directional information, while the polynomial-based method tends to miss or dampen small-scale fluctuations. For the purposes of this paper, we will only be conducting spectral analysis of TEC gradients computed using the polynomial-based method. The following section will outline the procedure used to spectrally analyze the time series of TEC gradients derived through the polynomial-based method.
\section{Spectral Analysis}
\label{sec:spectalanalysis}

\subsection{Methodology}
\label{sec:methodology}
The Fourier-based approach using TEC gradient measurements obtained from the polynomial method is used to detect and analyze individual or groups of waves passing over the array. This method was previously utilized by \citeA{Helm2012RaSc...47.0L02H_spectral}, to account for wavefront distortions and detect multiple wavefronts simultaneously. \par
The primary objective is to improve the array's spatial coverage by considering the source apparent position and movements of ionospheric disturbances, which have different speeds, wavelengths, and directions. To estimate the properties of each structure, the TEC fluctuations are approximated as a combination of multiple oscillating modes, each having the specific form
\begin{linenomath*}
\begin{eqnarray}
    {\rm TEC}(t) = \tau_\omega \ exp[i (\omega t-k_{x}x-k_{y}y)]
\end{eqnarray}
\end{linenomath*}
where $\tau_\omega$ is the amplitude which varies perpendicular to wavefront distortions, $\omega$ is the oscillation frequency and $k_x$ \& $k_y$ are the wave number for position $x$ (North-South) \& $y$ (East-West) respectively. For a single Fourier mode, the Fourier transforms of the partial derivatives with respect to $x$ and $y$ are:
\begin{linenomath*}
\begin{eqnarray}
    \label{eq:FT_theory}
    D_x(\omega;x,y) = \left[\frac{\partial \tau_\omega}{\partial x} - ik_x\tau_\omega \right] \times exp[-i(k_xx+k_yy)] \nonumber \\
    D_y(\omega;x,y) = \left[\frac{\partial \tau_\omega}{\partial y} - ik_y\tau_\omega \right] \times exp[-i(k_xx+k_yy)] 
\end{eqnarray}
\end{linenomath*}

\citeA{Mangla2022MNRAS.513..964M} employed polynomial series method to estimate TEC in the plane transverse to the line of sight. This method assumes that TEC can be represented by a two-dimensional, third-order polynomial, such that the difference in TEC between two antennas $i$ and $j$ can be represented as

\begin{linenomath*}
\begin{eqnarray}
    \label{eq:dpoly_2nd}
    \Delta \mbox{TEC}_{ij} \! &=& \! p_0\,(x_i-x_j) + p_1\,(y_i-y_j) + p_2\,(x_i^2-x_j^2) \nonumber \\ 
    &+& \! p_3\,(y_i^2-y_j^2) + p_4\,(x_i\,y_i-x_j\,y_j) \nonumber \\
    &+& \! p_5\,(x_i^3-x_j^3) + p_6\,(y_i^3-y_j^3) + p_7\,(x_i^2\,y_i-x_j^2\,y_j) \nonumber \\
    &+& \! p_8\,(x_i\,y_i^2-x_j\,y_j^2)
\end{eqnarray}
\end{linenomath*}

where x and y are the antenna positions projected onto the transverse plane in the north-south and east-west directions, respectively. At any given time, this method has the capability to replicate most of the structure observed.  Subsequently, the polynomial coefficients time series can then be Fourier transformed to decompose the transverse gradient in TEC into the Fourier modes \cite{Helmboldt2020a}, such that

\begin{linenomath*}
\begin{eqnarray}
    \label{eq:FT_observe}
    D_x(\omega;x,y) &=& P_0(\omega) + 2P_2(\omega)x + P_4(\omega)y + 3P_5(\omega)x^2 \nonumber \\
    &+&  2P_7(\omega)xy + P_8(\omega)y^2 \nonumber \\
    D_y(\omega;x,y) &=& P_1(\omega) + 2P_3(\omega)y + P_4(\omega)x + 3P_6(\omega)y^2 \nonumber \\
    &+&  P_7(\omega)x^2 + 2P_8(\omega)xy
\end{eqnarray}
\end{linenomath*}
where $D_x$ and $D_y$ represent the Fourier transforms of the north-south and east-west components of the TEC gradient, respectively, while $P_n(\omega)$ corresponds to the Fourier transform of $p_n(t)$. By expanding equation \ref{eq:FT_theory} using Taylor series and comparing the variables ($x$, $y$, $x^2$, $y^2$, $xy$) with equation \ref{eq:FT_observe} for respective $D_x$ and $D_y$, the wave number vector components ($k_x$ and $k_y$) for each mode can be approximated as
\begin{linenomath*}
\begin{eqnarray}
    \label{eq:wavenumbers}
    k_x(\omega) \simeq - Im \left\{ \frac{2P_2}{P_0}(\omega) \right\}  \simeq - Im \left\{ \frac{P_4}{P_1}(\omega) \right\} \nonumber \\
    k_y(\omega) \simeq - Im \left\{ \frac{2P_3}{P_1}(\omega) \right\}  \simeq - Im \left\{ \frac{P_4}{P_0}(\omega) \right\}
\end{eqnarray}
\end{linenomath*}

Since there are two estimators for both $k_x$ and $k_y$, a weighted average is utilised, with the weights being $|P_0|^2$ and $|P_1|^2$. 
It is to be noted that in the above equation $P_5$ to $P_8$ are not considered as they will induced higher-order effects (explained in Appendix \ref{appendix_a}). We also estimated the spectral power, wave speed and azimuth (propagating wave direction) respectively as 

\begin{linenomath*}
\begin{eqnarray}
    \label{eq:Power}
    \mathcal{P} \equiv \big|\mathcal{F} \{\Delta \mbox{TEC}\}\big|^{2}
\end{eqnarray}
\begin{eqnarray}
    \label{eq:velocity}
    V \simeq \frac{\omega}{\sqrt{k^{2}_x + k^{2}_y}}
\end{eqnarray}
\begin{eqnarray}
    \label{eq:azimuth}
    Az \simeq tan^{-1} \left(\frac{k_y}{k_x}\right)
\end{eqnarray}
\end{linenomath*}
\subsection{Derived Wave Properties from GMRT data}
\label{sec:waveproperties_poly}
\begin{figure}
    \centering
    \includegraphics[width=0.83\textwidth]{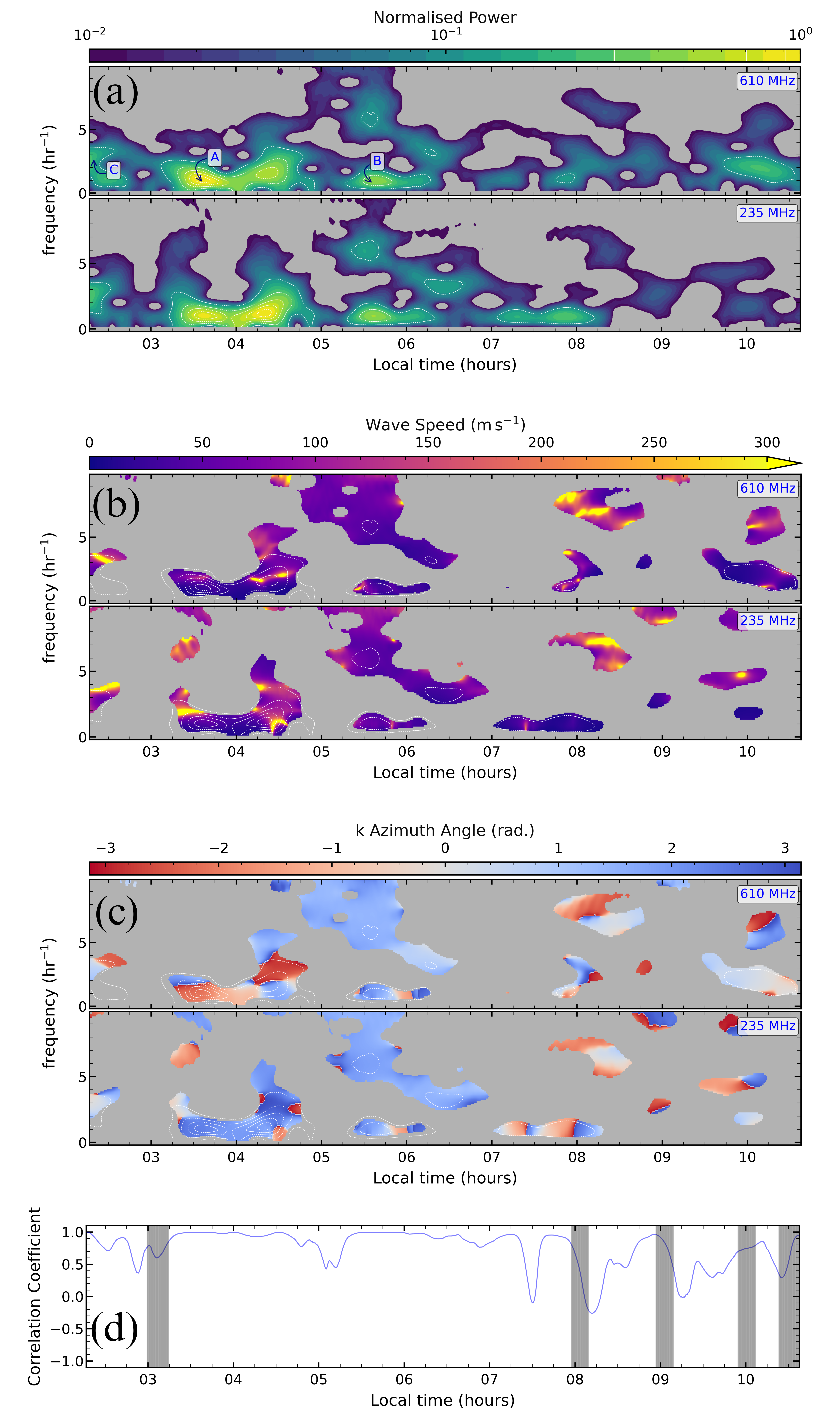}
    \caption{Spectral properties at 235 and 610\,MHz.  In all three panels (a, b, and c), the white contours are of spectral power. (a) The normalised power in the direction of the wave number vector as a function of temporal frequency (y-axis) and local time (x-axis). Similar structures are present in both bands, in addition, more smaller structures are visible in 235\,MHz band compared to 610\,MHz band. (b) The wave speed of each Fourier mode with power significantly larger than background noise.(c) the propagation direction or azimuth angle (measured clockwise from North through East) of wave number vector, using the same masking as in the (b) panel. (d) Using power spectra at 235 and 610\,MHz, Pearson correlation coefficient is computed, which is calculated independently for each time step.}
    \label{fig:fig1}
\end{figure}
Polynomial-based TEC gradient fits are used to compute the power as a function of time and frequency at two separate frequency bands (235 and 610\,MHz) by performing discrete Fourier transforms (DFTs) on the time series of polynomial coefficients. To enable the use of DFTs, which work best with evenly sampled data, the missing time steps of 10\,min between each scan were filled with zeros (a process called `zero-padding'). The DFTs are performed with a sliding window (`Hamming window') of one hour for frequencies up to $10\,\rm hr^{-1}$ (or, period $>$ 6.0 min), as the same one hour is used to de-trend the \dtec data \cite{Mangla2022MNRAS.513..964M}. Choosing Hamming window will diminish the ringing effects that might form due to the zero-padding. The power is computed for a period of 30\,min before and after the observing run. \par
Therefore, the DFTs of each polynomial coefficient is used to calculate the power as a function of temporal frequency and time for the two bands. Fig. \ref{fig:fig1}(a) displays the normalised power for bands 235 and 610\,MHz. Although random fluctuations are commonly observed,  significant wave detections above the background are observed in multiple cases. Most of these waves are observable at frequencies ranging from 0.5 to 4 hr$^{-1}$, or periods between 15 and 120\,min, which is typical behaviour of MSTIDs. To isolate such detections, a mask is produced by determining the median and median absolute deviation (MAD) within elliptical `annuli' around each pixel with a size of 1.5\,hr in local time and 2\,hr$^{-1}$ in frequency. A pixel exceeding two times the MAD over median for its annulus is considered a detection above the background. \par
The wave's speed and propagation direction or azimuth angle (measured clockwise from North through East) are calculated as a function of local time and temporal frequency using equations \ref{eq:velocity} and \ref{eq:azimuth}. The same mask is used to display both the wave speed and azimuth angle for the detection, which were shown in Figs. \ref{fig:fig1}(b) and \ref{fig:fig1}(c) respectively for 235 and 610\,MHz bands. 

\begin{sidewaystable}
        \caption{Wave parameters derived from GMRT measurements at 235 and 610 MHz}
    \centering
    \begin{tabular}{cccccccc}
    \hline
    \multicolumn{1}{c}{Events}  & \multicolumn{1}{c}{Period}  & \multicolumn{2}{c}{Wave speed} & \multicolumn{2}{c}{Azimuth} & \multicolumn{2}{c}{Wavelength} \\
    \multicolumn{1}{c}{}        & \multicolumn{1}{c}{}        & \multicolumn{2}{c}{(m/s)}    & \multicolumn{2}{c}{(degree)} & \multicolumn{2}{c}{(km)}\\
    \hline
        & & 610\,MHz & 235\,MHz & 610\,MHz & 235\,MHz & 610\,MHz & 235\,MHz\\
    \hline
    \vspace{3mm}
        A & 1\,hr 10\,min & 46.55$^{\pm5.16}$ & 48.02$^{\pm6.54}$ & -81.78$^{\pm0.67}$ (W) & 135.09$^{\pm0.59}$ (SE) & 195.51$^{\pm21.67}$ & 201.68$^{\pm27.48}$ \\ 
    \vspace{3mm}
        B & 1\,hr 33\,min & 54.91$^{\pm3.45}$ & 56.78$^{\pm5.23}$ & 101.52$^{\pm3.36}$ (E)  & 103.13$^{\pm6.83}$ (E) & 263.57$^{\pm16.56}$ & 272.54$^{\pm25.10}$ \\
    \vspace{1mm}
        C & 24\,min & 95.47$^{\pm9.20}$ & 106.20$^{\pm8.02}$ & 28.07$^{\pm10.42}$ (N) & 13.18$^{\pm1.49}$ (N) & 137.19$^{\pm13.25}$ & 152.93$^{\pm11.55}$ \\
    \hline
    \end{tabular}
    \label{tab:table1}
\end{sidewaystable}
To determine whether the wave structures in the two frequency bands are similar, we computed the Pearson correlation coefficients. Fig. \ref{fig:fig1}(d) shows the correlation coefficient computed independently for every time stamp, with the grey vertical lines indicating the time stamps where zero padding is used. The normalised power is highly correlated, especially for the longer duration of scan (scan no. 4; nearly 5\,hr), emphasizing the importance of continuous data for spectral analysis in ionospheric studies. There are more detections visible in the 235\,MHz band compared to the 610\,MHz band, indicating the dependence of refraction on the observable frequency. The correlation coefficient time series is partially correlated or decorrelated where the scan is very short or the detection is less above the background. A substantial number of detections, including small structures, are highly correlated between local times of 05:00 and 06:00 hours, suggesting that the Earth's ionosphere varies before dawn. Additionally, several small-scale structures are detected in the direction of a pre-existing MSTID, which is towards the eastward. \par
Based on the highly correlated coefficient observed in Fig. \ref{fig:fig1}(d), we have selected three events indicated by \texttt{A, B} and \texttt{C} in the Fig. \ref{fig:fig1}(a). We have calculated the properties of the events using the method described in Section \ref{sec:methodology} and presented the results in Table \ref{tab:table1}.  Events \texttt{B} and \texttt{C} show agreement in wave speed and azimuth angle, while event \texttt{A} only partially agrees. The wave speed in the two band for event \texttt{A} is approximately the same, but the propagation  direction is out of phase by approximately 180$^{\circ}$, which may be due to scintillation observed in the higher order terms of polynomial fitting at 235\,MHz data \cite{Mangla2022JoAA}.
\subsubsection{Mean Power Spectra}
\label{sec:meanpowerspectra_poly}
\begin{figure}
    \centering
    \includegraphics[width=\textwidth]{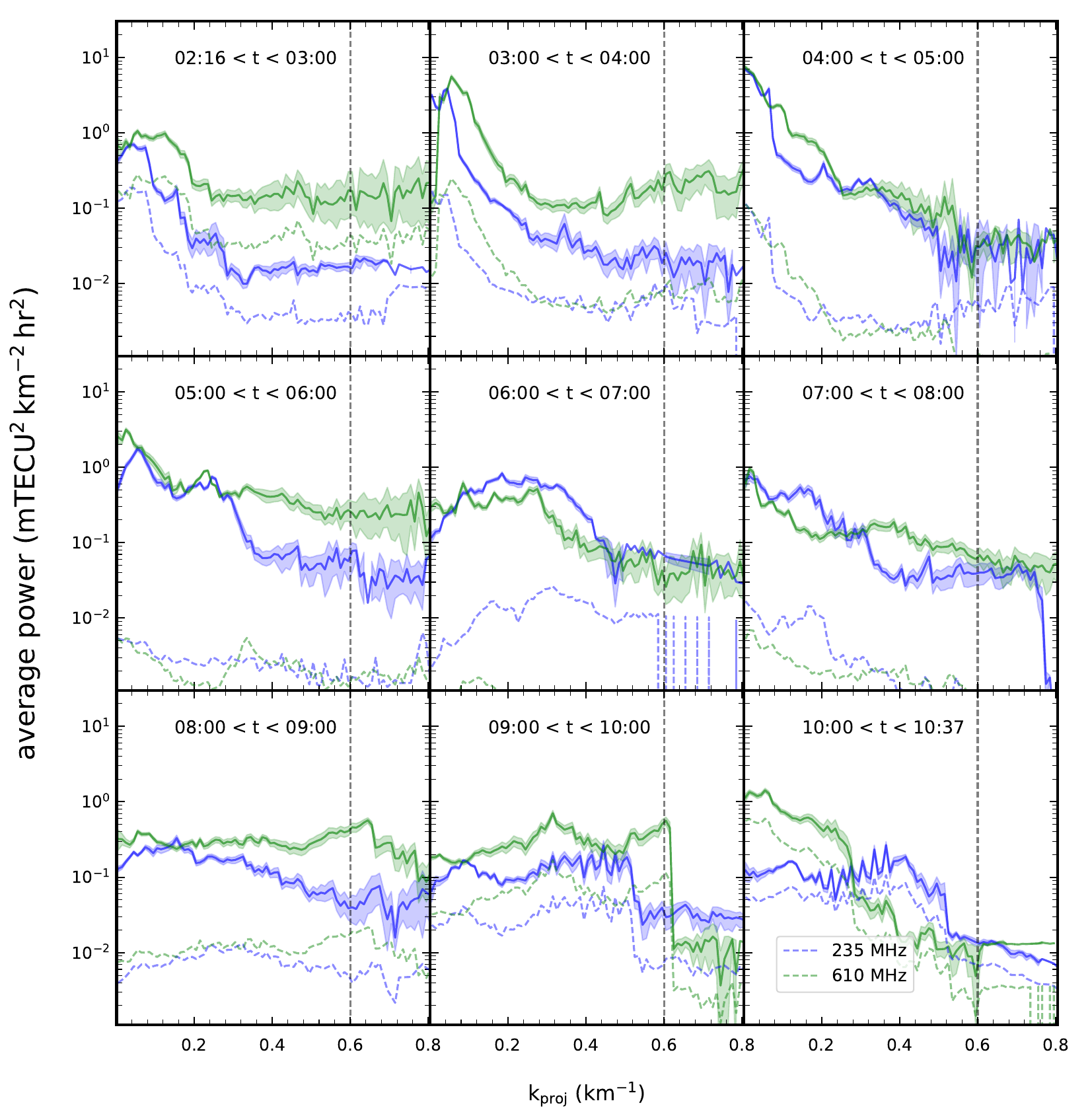}
    \caption{Within one-hour bins, the average power for 610 and 235\,MHz as a function of wave number, k, for the data displayed in Fig. \ref{fig:fig1}. Noise-equivalent spectra are also plotted in dashed curves but with the same color as average power spectra of corresponding band.}
    \label{fig:fig2}
\end{figure}
A more statistical description of the observed set of TEC variations may be produced using the polynomial-based methodology. While Figure \ref{fig:fig1} provide useful information for locating and evaluating specific instances of waves or groups of waves, we presented average power as a function of projected wave number inside bins of wave number (k) for each of the one-hour blocks of time in Figure \ref{fig:fig2}. The shaded part represents the average power, and the solid line signifies the median in that region. Additionally, we included the corresponding `noise-equivalent' spectra (dashed lines) for 235 and 610\,MHz in blue and green, respectively. The noise-equivalent spectra are computed using the following method:

\begin{enumerate}
    \item First, by performing the 2D polynomial fitting on the \dtec measurements on each of the bands and mean \dtec measurement. This process is very well explained in \citeA{Mangla2022MNRAS.513..964M}.
    \item Then by performing the DFTs on the difference between the individual fits and fit to the final mean \dtec time series.
    \item And lastly, to get the noise-equivalent spectra, the same analysis is performed on the residual DFTs which is detailed in Sect. \ref{sec:methodology}.
\end{enumerate}
In Figure \ref{fig:fig2}, we can see the resulting `noise-equivalent' spectra. These spectra are the average spectra computed within each one-hour block. The average power spectra (shaded region) intersect with the noise-equivalent spectra of that band, which is around  $\rm \sim 0.6\,km^{-1}$ (dashed grey line) and is clearly seen in the figure corresponding to panels 03:00 $<$ 04:00 and 06:00 $<$ 07:00. This intersection suggests that the method is capable of detecting substantial power of structures that are smaller than half the size of the array, approximately 10\,km or $\rm k\,\sim 0.6\,km^{-1}$. Many structures are also visible in Fig. \ref{fig:fig1}(a), which can be identified as noticeable bumps above the background at wave numbers ranging from $\rm 0.015\,-\,0.12\,km^{-1}$, which corresponds to wavelengths of about 50 to 500\,km in the majority of the panels and consistent with the known properties of MSTIDs.
\setlength{\tabcolsep}{5pt}
\renewcommand{\arraystretch}{1.0} % Default 
\begin{sidewaystable}
    \footnotesize
    \centering
    \caption{Summary of Observed Phenomena at different geophysical location and with different/distinct instruments during different seasons}
    \begin{tabular}{lccccccc}
        \toprule
          Location        & Instruments & Period    & Wavelength& Wave speed & & Direction & References  \\
          &       &   (min)   &   (km)    &    (m$/$s)       & &    & or comments \\
          \midrule[.1em]
          &          & 10-20 & 100-220     &  50-200   & &   Northeast/Eastward (During Dusk) & \\ 
          \multirow{8}{4em}{\centering Low-Latitude} & GMRT & 60-90 & 150-250 & 40-100 & \multirow{1}{4em}{\textit{Summer}:} &  Westward (nighttime) & this work \\
          &          & $>$90 & 250-350 & 50-100 & & Eastward (During Dusk) & \\
          \cline{2-8}
          & \multirow{6}{4em}{\centering GPS} & \multirow{6}{4em}{\centering 15-30} & \multirow{6}{4em}{\centering 50-300} & \multirow{6}{4em}{\centering 50-250} & \multirow{2}{4em}{\textit{Fall}:} & Southeast (daytime) & \multirow{6}{10em}{\centering \cite{Hern2012RaSc...47.0K05H}} \\
          & & & & & & Northeast/Westward (nighttime) & \\
          & & & & & \multirow{2}{4em}{\textit{Winter}:} & Southeast (daytime) & \\
          & & & & & & Westward (nighttime) & \\
          & & & & & \multirow{1}{4em}{\textit{Spring}:} & Westward (nighttime) & \\
          & & & & & \multirow{1}{4em}{\textit{Summer}:} & Northeast/westward (nighttime) & \\
          \cline{2-8}
          & \multirow{2}{4em}{\centering Airglow Imager} & 30-90 & 100-400 & 30-120 & \multirow{1}{4em}{\textit{all}:} & \multirow{2}{10em}{\centering Southwestward (nighttime)} & \cite{Narayanan2014JGRA..119.9268L} \\
          &  & 15-20 & 100-175 & 130-140 & \multirow{1}{4em}{\textit{Winter}:} & & \cite{Sivakandan2019JGRA..124..749S} \\
          \hline
         \multirow{17}{4em}{\centering Mid-Latitude} & \multirow{3}{3em}{\centering VLA} & 15-50 & 160-300 & 110-170 & \multirow{3}{4em}{\textit{Summer}:} & Southwest & \multirow{1}{10em}{\centering \cite{Dymond...2010RS004535}} \\
         & & \multirow{2}{3em}{\centering $>$20} & & \multirow{2}{5em}{\centering 100-200} & & West/Northwest (Before Midnight) &  \multirow{2}{10em}{\centering \cite{Helm2012RaSc...47.0L02H_spectral}} \\
         & & & & & & Northeast (After midnight) & \\
         \cline{2-8}
         & \multirow{3}{4em}{\centering LOFAR} & & & 200 & \multirow{3}{4em}{\textit{Fall}:} & Southwest to Northeast &  \cite{gasp18} \\
         &  & & 200-700 & 20-40 & & Northwest to Southeast &  \multirow{1}{10em}{\centering \cite{Rich2020JSWSC..10...10F}} \\
         & & $\sim$11& 60-70 & $\sim$100 & & Northeast to Southwest & two TIDs (nighttime) \\
         \cline{2-8}
         & MWA & 60 & 700 & 200 & \multirow{1}{4em}{\textit{Spring}:} & Northeast & \cite{Loi2016JGRA..121.1569L} \\
         \cline{2-8}
         & \multirow{10}{4em}{\centering GPS} & \multirow{8}{4em}{\centering 15-27} & \multirow{8}{4em}{\centering 50-250} & \multirow{8}{4em}{\centering 50-250} & \multirow{2}{4em}{\textit{Fall}:} & Southeast (daytime) & \multirow{8}{10em}{\centering \cite{MSTID_gps,Hern2012RaSc...47.0K05H,Otsuka2013AnGeo..31..163O}} \\
         & & & & & & Northwest/Westward (nighttime) & \\
         & & & & & \multirow{2}{4em}{\textit{Winter}:} & Southeast (daytime) & \\
         & & & & & & Westward (nighttime) & \\
         & & & & & \multirow{2}{4em}{\textit{Spring}:} & Southeast/Eastward (daytime) & \\
         & & & & & & Westward (nighttime) & \\
         & & & & & \multirow{2}{4em}{\textit{Summer}:} & Eastward (daytime) & \\
         & & & & & & Westward (nighttime) & \\
         & & \multirow{2}{4em}{\centering 10-60} & 300-1000 & 100-200 & \multirow{1}{4em}{\textit{Winter}:} & Southeast (daytime)  & \multirow{2}{8em}{\centering \cite{Tsugawa2007GeoRL..3422101T}} \\
         & & &  200-500 & 100-150 & \multirow{1}{4em}{\textit{Summer}:} & Southwest (nighttime) & \\          
         \cline{2-8}
         & \multirow{2}{4em}{\centering Airglow Imager} & \multirow{2}{4em}{\centering 30-90} & \multirow{2}{4em}{\centering 100-300}& \multirow{2}{4em}{\centering 50-100} & \multirow{2}{4em}{\textit{all}:} & \multirow{2}{10em}{\centering Southwest (nighttime)} & \multirow{2}{8em}{\centering \cite{Shiokawa2003JGRA..108.1052S}} \\
         & & & & & & & \\
         \hline
         \multirow{7}{4em}{\centering High-Latitude} & \multirow{7}{4em}{\centering GPS} & \multirow{7}{4em}{\centering 10-25} & \multirow{7}{4em}{\centering 100-300} & \multirow{7}{4em}{\centering 100-300} & \multirow{2}{4em}{\textit{Fall}:} & Southeast (daytime) & \multirow{6}{10em}{\centering \cite{Hern2012RaSc...47.0K05H}} \\
          & & & & & & Northeast/Southwest (nighttime) & \\
          & & & & & \multirow{2}{4em}{\textit{Winter}:} & Southeast (daytime) & \\
          & & & & & & Northward/Westward (nighttime) & \\
          & & & & & \multirow{1}{4em}{\textit{Spring}:} & Southwest (nighttime) & \\
          & & & & & \multirow{2}{4em}{\textit{Summer}:} & Northwest (nighttime) & \\
          & & & & & & Southeast (dusktime) & \\
         \bottomrule             
    \end{tabular}
    \label{tab:comparison-interferomters}
\end{sidewaystable}

\section{Results and Conclusions}
\label{sec:conclusion}

Our work showed that the GMRT instrument can efficiently study ionospheric fluctuations using a long observation of a bright radio source. For the aforementioned observations, the $1\sigma$ uncertainty in the \dtec measurements is 1\,mTECU, in both the 235 and 610\,MHz bands [see \citeA{Mangla2022JoAA} and \citeA{Mangla2022MNRAS.513..964M}, respectively]. This level of sensitivity achieved by the GMRT is ten time better than that with GNSS measurements \cite{MSTID_gps}. \par
Spectral analysis technique can detect multiple MSTIDs simultaneously, and the wave structures in both bands are similar, as shown in Fig. \ref{fig:fig1}(a), with high correlation confirmed by the Pearson correlation coefficient computed at each time step. The normalized power is highly correlated, particularly in scan no. 4, which lasted over 4 hours. Thus, ionospheric studies using spectral analysis would provide better results, if continuous observation of bright radio source is taken. Furthermore, Fig. \ref{fig:fig1}(a) suggest possible MSTIDs candidates with periods between 10 and 30\,min, or longer, and estimated speeds between 50 and 150 m$\rm s^{-1}$, which is similar to nighttime MSTIDs in the Northern Hemisphere \cite{CROWLEY2018555_classification_TIDs}. During summer, nighttime MSTIDs primarily propagate southwestward \cite{Tsugawa2007GeoRL..3422101T, Otsuka2013AnGeo..31..163O}, likely generated by the ionosphere's electrodynamics. Mid-latitude daytime MSTIDs mostly propagate southeastward and are likely generated by acoustic gravity waves \cite{Kotake2007EP&S...59...95K}. \par

This work also identified a group of waves between local hours 05:00 and 06:00, indicating unexpected ionospheric changes during sunrise, similar to those anticipated during sunset \cite{MSTID_gps}. Additionally, polynomial-based methods can detect structures as small as 10\,km , showing the GMRT's ability to detect ionospheric variations at such small wavelengths. Among the many detected waves present intermittently throughout the night and turbulent fluctuations seen at all times, the waves appear to change direction after sunrise. This phenomenon is visible in both frequency bands for waves with oscillation frequencies between $\rm 3\,hr^{-1}$ and $\rm 9\,hr^{-1}$. \par
Our works key findings are summarized in table \ref{tab:comparison-interferomters}, along with other studies that have examined the MSTIDs propagation characteristics using data with different instruments, methods, locations, and time periods. We see that the wave speed and wavelengths obtained with the GMRT is comparable with the observed trend with GPS studies \cite{Hern2012RaSc...47.0K05H} at similar geophysical latitudes. Also, to explore further, we have also chosen three events on the basis of correlation-coefficient, that are shown in Fig. \ref{fig:fig1}(a) and named them as events \texttt{A, B} and \texttt{C}. The properties of these events are as follows:

\subsection{Event A}
During the entire observation run, a dominating pattern with a period of 1\,hr 10\,min and propagating at around 50\,m$\rm s^{-1}$ is briefly observed in both bands, as shown in Fig. \ref{fig:fig1}(d), near 03:40 Local time (UTC\,+\,05:30). This pattern exhibits MSTID-like properties and resembling what has been observed during nighttime summer season at low-latitudes using GPS studies \cite{Hern2012RaSc...47.0K05H}. However, during the same time, ionospheric scintillation was likely observed at 235\,MHz in the higher-order terms of polynomial fitting \cite{Mangla2022JoAA}, which highlights a limitation of this method (to determining the propagation speed accurately), as phase measurements for radio interferometers are sensitive to induced phase errors from ionospheric scintillation or other sources. Furthermore, the Fresnel scale at an altitude of 300\,km for 235\,MHz is about 620\,m. These observations are therefore sensitive to structures at these scales. \citeA{Helmboldt_2022} also noted a similar correlation between TID activity and scintillation at 35\,MHz (Fresnel scale $\sim$ 1.5\,km). 
\subsection{Event B}
During dusk-time (05:15 to 06:15 Local time), eastward directed wave is observed along with many small waves in both the bands. It is unusual to observe ionospheric wave during sunrise as the ionosphere is in the state of transition and the dynamics are more complex. However, this explains the formation of many small structures detected in both the bands in the similar direction. It may also be the case that these smaller waves are present inside the detected ionospheric structure propagating in the same direction. Furthermore, from Fig. \ref{fig:fig2}, one can also notice the two to three orders of magnitude less noise in the panels displaying the duration of 05:00 $<$ 07:00 local time. \par
However, it may be possible that these ionospheric disturbances are the cases of equatorial spread\,$F$ (ESF) which occurs the daytime (typically around sunrise and sunset) at low-latitudes \cite{TsunodaJA086iA05p03610, Sultan1996JGR...10126875S, Fejer1999JGR...10419859F, Pacheco2011JGRA..11611329P, Mrak2021}. Furthermore, observed ionospheric structure has the period and wavelength of 1\,hr 33\,min and $\sim$\,300\,km respectively propagating with a speed of somewhere between 50 to 60\,m$\rm s^{-1}$, which falls within the typical range of ESF. A more comprehensive, statistical analysis of these phenomena appears to be necessary. Besides, it is important to note that ESF is a complex phenomenon and is still not fully understood, but it is known to have a significant impact on the ionosphere and the propagation of radio signals, especially in the equatorial regions. The study of ESF is important for understanding the dynamics of the ionosphere, and for mitigating the impact of ionospheric disturbances on communication and navigation systems.
\subsection{Event C}
We observed northward moving MSTID briefly during the nighttime (02:15 Local time) with a period of 24\,min and wave speed of about 100\,m$\rm s^{-1}$ in both bands. Nighttime northward propagating MSTIDs are rare events in the northern hemisphere and \citeA{Ichihara201342} suggested that the northward propagation of MSTIDs is caused by gravity waves. Similar event have been observed by \citeA{Bhat2021AdSpR..68.3806B} using the Airglow imager in the Indian logitude sector. \par
Furthermore, \citeA{Shiokawa_2005JA011406} discovered quasiperiodic southward-moving MSTIDs during nighttime, likely caused by gravity waves in the thermosphere. Similar studies by \cite{Fukushima2012JGRA..11710324F} also observed southward propagation of nighttime MSTIDs and hypothesized that deep convection in the troposphere generates gravity waves that contribute to the formation of MSTIDs. Both of these studies were conducted using 630.0\,nm airglow observations in the Kototabang region of Indonesia, which is located in the equatorial latitudes, but in the southern hemisphere. These studies is consistent with our work, which shows that the propagation direction is toward poles. \par

The northward moving TIDs during night time (Event C), and eastward moving TIDs during the sunrise time (Event B), may be attributed to gravity waves (GWs) that have been filtered by the wind. Gravity waves that have undergone wind filtering tend to propagate more easily in the opposite direction of the wind, while facing difficulties propagating in the same direction as the wind. Based on the Horizontal Wind Model (HWM) climatology \cite{Drob_2015}, it is indicated that the winds within the F-region would have been moving south by southwest around midnight and subsequently transitioning to predominantly westward flow by 05:00 local time. Therefore, it is plausible to suggest that the detected TIDs could be attributed to wind-filtered GWs. Similar phenomenon is observed by \citeA{Mukherjee_2010} during summer time, using an All-sky imager in Indian longitudinal region. \par
This work is just the beginning of using GMRT to detect MSTIDs. More work is needed to automate the detection process and improve localization and propagation estimation. Simultaneous observations from GPS stations would be particularly valuable in unraveling the microphysical nature of TIDs, as GNSS data primarily captures large- to mid-scale behavior. In contrast, radio interferometers can provide insights into the small-scale behavior, bridging the gap between these two scales and enabling a more comprehensive understanding of the microphysical processes driving TIDs formation and the associated directional behaviors. \par
Additionally, due to greater spacial and temporal resolution of a radio telescopes, they can provide high-quality measurements of ionospheric parameters that can be used to develop more accurate and sophisticated models of the ionosphere, which will further help to correct for ionospheric delays in Interferometric synthetic aperture radar (InSAR) and geodetic instruments data. This is important because ionospheric delays can cause errors in InSAR measurements of ground deformation, particularly for long-wavelength signals. In addition, radio telescopes can provide information on ionospheric irregularities and scintillations, which can affect the propagation of radio signals and cause errors in navigation and communication systems. By combining data from InSAR and radio telescopes, researchers can improve their understanding of the ionosphere and its effects on geodetic and communication applications.

\section*{Open Research}
All the radio observation data used in this study are available in the GMRT Online Archive (\url{https://naps.ncra.tifr.res.in/goa/data/search}) with proposal code 22\_064. Solar and geomagnetic indices are obtained from the OMNIWeb services (\url{https://omniweb.gsfc.nasa.gov/ow.html}).

\acknowledgments
We thank the staff of the GMRT who have made these observations possible. GMRT is run by the National Centre for Radio Astrophysics of the Tata Institute of Fundamental Research. SM is grateful for the financial assistance from the University Grants Commission. We thank the two anonymous reviewers whose comments/suggestions helped improve and clarify this manuscript.   \par
This work made use of \texttt{ASTROPY}, a community-developed core Python package for Astronomy \cite{astropy:2013, astropy:2018}, \texttt{NUMPY} \cite{Numpy2020array}, \texttt{SCIPY} \cite{SciPy-NMeth2020}. This research also made use of \texttt{MATPLOTLIB} \cite{matplotlib07} and \texttt{SEABORN} \cite{seaborn2021} open-source plotting packages for \texttt{PYTHON}.

\bibliography{agusample}

\begin{thebibliography}{}

\bibitem [\protect \citeauthoryear {%
{Appleton}%
}{%
{Appleton}%
}{%
{\protect \APACyear {1946}}%
}]{%
Appl1946Natur.157..691A}
\APACinsertmetastar {%
Appl1946Natur.157..691A}%
\begin{APACrefauthors}%
{Appleton}, E\BPBI V.%
\end{APACrefauthors}%
\unskip\
\newblock
\APACrefYearMonthDay{1946}{{\APACmonth{05}}}{}.
\newblock
{\BBOQ}\APACrefatitle {{Two Anomalies in the Ionosphere}} {{Two Anomalies in
  the Ionosphere}}.{\BBCQ}
\newblock
\APACjournalVolNumPages{Nature}{157}{3995}{691}.
\newblock
\begin{APACrefDOI} \doi{10.1038/157691a0} \end{APACrefDOI}
\PrintBackRefs{\CurrentBib}

\bibitem [\protect \citeauthoryear {%
{Astropy Collaboration}%
\ \protect \BOthers {.}}{%
{Astropy Collaboration}%
\ \protect \BOthers {.}}{%
{\protect \APACyear {2018}}%
}]{%
astropy:2018}
\APACinsertmetastar {%
astropy:2018}%
\begin{APACrefauthors}%
{Astropy Collaboration}%
, {Price-Whelan}, A\BPBI M.%
, {Sip{\H{o}}cz}, B\BPBI M.%
, {G{\"u}nther}, H\BPBI M.%
, {Lim}, P\BPBI L.%
, {Crawford}, S\BPBI M.%
\BDBL {}{Astropy Contributors}%
\end{APACrefauthors}%
\unskip\
\newblock
\APACrefYearMonthDay{2018}{{\APACmonth{09}}}{}.
\newblock
{\BBOQ}\APACrefatitle {{The Astropy Project: Building an Open-science Project
  and Status of the v2.0 Core Package}} {{The Astropy Project: Building an
  Open-science Project and Status of the v2.0 Core Package}}.{\BBCQ}
\newblock
\APACjournalVolNumPages{The Astronomical Journal}{156}{3}{123}.
\newblock
\begin{APACrefDOI} \doi{10.3847/1538-3881/aabc4f} \end{APACrefDOI}
\PrintBackRefs{\CurrentBib}

\bibitem [\protect \citeauthoryear {%
{Astropy Collaboration}%
\ \protect \BOthers {.}}{%
{Astropy Collaboration}%
\ \protect \BOthers {.}}{%
{\protect \APACyear {2013}}%
}]{%
astropy:2013}
\APACinsertmetastar {%
astropy:2013}%
\begin{APACrefauthors}%
{Astropy Collaboration}%
, {Robitaille}, T\BPBI P.%
, {Tollerud}, E\BPBI J.%
, {Greenfield}, P.%
, {Droettboom}, M.%
, {Bray}, E.%
\BDBL {}{Streicher}, O.%
\end{APACrefauthors}%
\unskip\
\newblock
\APACrefYearMonthDay{2013}{{\APACmonth{10}}}{}.
\newblock
{\BBOQ}\APACrefatitle {{Astropy: A community Python package for astronomy}}
  {{Astropy: A community Python package for astronomy}}.{\BBCQ}
\newblock
\APACjournalVolNumPages{Astronomy and Astrophysics}{558}{}{A33}.
\newblock
\begin{APACrefDOI} \doi{10.1051/0004-6361/201322068} \end{APACrefDOI}
\PrintBackRefs{\CurrentBib}

\bibitem [\protect \citeauthoryear {%
{Bhat}%
, {Ganaie}%
, {Ramkumar}%
\BCBL {}\ \BBA {} {Malik}%
}{%
{Bhat}%
\ \protect \BOthers {.}}{%
{\protect \APACyear {2021}}%
}]{%
Bhat2021AdSpR..68.3806B}
\APACinsertmetastar {%
Bhat2021AdSpR..68.3806B}%
\begin{APACrefauthors}%
{Bhat}, A\BPBI H.%
, {Ganaie}, B\BPBI A.%
, {Ramkumar}, T\BPBI K.%
\BCBL {}\ \BBA {} {Malik}, M\BPBI A.%
\end{APACrefauthors}%
\unskip\
\newblock
\APACrefYearMonthDay{2021}{{\APACmonth{11}}}{}.
\newblock
{\BBOQ}\APACrefatitle {{Northward propagation of medium scale traveling
  ionospheric disturbances over Srinagar, J and K India}} {{Northward
  propagation of medium scale traveling ionospheric disturbances over Srinagar,
  J and K India}}.{\BBCQ}
\newblock
\APACjournalVolNumPages{Advances in Space Research}{68}{9}{3806-3813}.
\newblock
\begin{APACrefDOI} \doi{10.1016/j.asr.2021.06.035} \end{APACrefDOI}
\PrintBackRefs{\CurrentBib}

\bibitem [\protect \citeauthoryear {%
Crowley%
\ \BBA {} Azeem%
}{%
Crowley%
\ \BBA {} Azeem%
}{%
{\protect \APACyear {2018}}%
}]{%
CROWLEY2018555_classification_TIDs}
\APACinsertmetastar {%
CROWLEY2018555_classification_TIDs}%
\begin{APACrefauthors}%
Crowley, G.%
\BCBT {}\ \BBA {} Azeem, I.%
\end{APACrefauthors}%
\unskip\
\newblock
\APACrefYearMonthDay{2018}{}{}.
\newblock
{\BBOQ}\APACrefatitle {Chapter 23 - Extreme Ionospheric Storms and Their
  Effects on GPS Systems} {Chapter 23 - extreme ionospheric storms and their
  effects on gps systems}.{\BBCQ}
\newblock
\BIn{} N.~Buzulukova\ (\BED), \APACrefbtitle {Extreme Events in Geospace}
  {Extreme events in geospace}\ (\BPG~555-586).
\newblock
\APACaddressPublisher{}{Elsevier}.
\newblock
\begin{APACrefURL}
  \url{https://www.sciencedirect.com/science/article/pii/B9780128127001000236}
  \end{APACrefURL}
\newblock
\begin{APACrefDOI} \doi{https://doi.org/10.1016/B978-0-12-812700-1.00023-6}
  \end{APACrefDOI}
\PrintBackRefs{\CurrentBib}

\bibitem [\protect \citeauthoryear {%
{de Gasperin, F.}%
, {Mevius, M.}%
, {Rafferty, D. A.}%
, {Intema, H. T.}%
\BCBL {}\ \BBA {} {Fallows, R. A.}%
}{%
{de Gasperin, F.}%
\ \protect \BOthers {.}}{%
{\protect \APACyear {2018}}%
}]{%
gasp18}
\APACinsertmetastar {%
gasp18}%
\begin{APACrefauthors}%
{de Gasperin, F.}%
, {Mevius, M.}%
, {Rafferty, D. A.}%
, {Intema, H. T.}%
\BCBL {}\ \BBA {} {Fallows, R. A.}%
\end{APACrefauthors}%
\unskip\
\newblock
\APACrefYearMonthDay{2018}{}{}.
\newblock
{\BBOQ}\APACrefatitle {The effect of the ionosphere on ultra-low-frequency
  radio-interferometric observations} {The effect of the ionosphere on
  ultra-low-frequency radio-interferometric observations}.{\BBCQ}
\newblock
\APACjournalVolNumPages{A\&A}{615}{}{A179}.
\newblock
\begin{APACrefURL} \url{https://doi.org/10.1051/0004-6361/201833012}
  \end{APACrefURL}
\newblock
\begin{APACrefDOI} \doi{10.1051/0004-6361/201833012} \end{APACrefDOI}
\PrintBackRefs{\CurrentBib}

\bibitem [\protect \citeauthoryear {%
{Drob}%
\ \protect \BOthers {.}}{%
{Drob}%
\ \protect \BOthers {.}}{%
{\protect \APACyear {2015}}%
}]{%
Drob_2015}
\APACinsertmetastar {%
Drob_2015}%
\begin{APACrefauthors}%
{Drob}, D\BPBI P.%
, {Emmert}, J\BPBI T.%
, {Meriwether}, J\BPBI W.%
, {Makela}, J\BPBI J.%
, {Doornbos}, E.%
, {Conde}, M.%
\BDBL {}{Klenzing}, J\BPBI H.%
\end{APACrefauthors}%
\unskip\
\newblock
\APACrefYearMonthDay{2015}{{\APACmonth{07}}}{}.
\newblock
{\BBOQ}\APACrefatitle {{An update to the Horizontal Wind Model (HWM): The quiet
  time thermosphere}} {{An update to the Horizontal Wind Model (HWM): The quiet
  time thermosphere}}.{\BBCQ}
\newblock
\APACjournalVolNumPages{Earth and Space Science}{2}{7}{301-319}.
\newblock
\begin{APACrefDOI} \doi{10.1002/2014EA000089} \end{APACrefDOI}
\PrintBackRefs{\CurrentBib}

\bibitem [\protect \citeauthoryear {%
Dymond%
\ \protect \BOthers {.}}{%
Dymond%
\ \protect \BOthers {.}}{%
{\protect \APACyear {2011}}%
}]{%
Dymond...2010RS004535}
\APACinsertmetastar {%
Dymond...2010RS004535}%
\begin{APACrefauthors}%
Dymond, K\BPBI F.%
, Watts, C.%
, Coker, C.%
, Budzien, S\BPBI A.%
, Bernhardt, P\BPBI A.%
, Kassim, N.%
\BDBL {}Datta, A.%
\end{APACrefauthors}%
\unskip\
\newblock
\APACrefYearMonthDay{2011}{}{}.
\newblock
{\BBOQ}\APACrefatitle {A medium-scale traveling ionospheric disturbance
  observed from the ground and from space} {A medium-scale traveling
  ionospheric disturbance observed from the ground and from space}.{\BBCQ}
\newblock
\APACjournalVolNumPages{Radio Science}{46}{5}{}.
\newblock
\begin{APACrefURL}
  \url{https://agupubs.onlinelibrary.wiley.com/doi/abs/10.1029/2010RS004535}
  \end{APACrefURL}
\newblock
\begin{APACrefDOI} \doi{https://doi.org/10.1029/2010RS004535} \end{APACrefDOI}
\PrintBackRefs{\CurrentBib}

\bibitem [\protect \citeauthoryear {%
{Fallows}%
\ \protect \BOthers {.}}{%
{Fallows}%
\ \protect \BOthers {.}}{%
{\protect \APACyear {2020}}%
}]{%
Rich2020JSWSC..10...10F}
\APACinsertmetastar {%
Rich2020JSWSC..10...10F}%
\begin{APACrefauthors}%
{Fallows}, R\BPBI A.%
, {Forte}, B.%
, {Astin}, I.%
, {Allbrook}, T.%
, {Arnold}, A.%
, {Wood}, A.%
\BDBL {}{Zucca}, P.%
\end{APACrefauthors}%
\unskip\
\newblock
\APACrefYearMonthDay{2020}{{\APACmonth{02}}}{}.
\newblock
{\BBOQ}\APACrefatitle {{A LOFAR observation of ionospheric scintillation from
  two simultaneous travelling ionospheric disturbances}} {{A LOFAR observation
  of ionospheric scintillation from two simultaneous travelling ionospheric
  disturbances}}.{\BBCQ}
\newblock
\APACjournalVolNumPages{Journal of Space Weather and Space Climate}{10}{}{10}.
\newblock
\begin{APACrefDOI} \doi{10.1051/swsc/2020010} \end{APACrefDOI}
\PrintBackRefs{\CurrentBib}

\bibitem [\protect \citeauthoryear {%
{Fejer}%
, {Scherliess}%
\BCBL {}\ \BBA {} {de Paula}%
}{%
{Fejer}%
\ \protect \BOthers {.}}{%
{\protect \APACyear {1999}}%
}]{%
Fejer1999JGR...10419859F}
\APACinsertmetastar {%
Fejer1999JGR...10419859F}%
\begin{APACrefauthors}%
{Fejer}, B\BPBI G.%
, {Scherliess}, L.%
\BCBL {}\ \BBA {} {de Paula}, E\BPBI R.%
\end{APACrefauthors}%
\unskip\
\newblock
\APACrefYearMonthDay{1999}{{\APACmonth{09}}}{}.
\newblock
{\BBOQ}\APACrefatitle {{Effects of the vertical plasma drift velocity on the
  generation and evolution of equatorial spread F}} {{Effects of the vertical
  plasma drift velocity on the generation and evolution of equatorial spread
  F}}.{\BBCQ}
\newblock
\APACjournalVolNumPages{Journal of Geophysical Research: Space
  Physics}{104}{A9}{19859-19870}.
\newblock
\begin{APACrefDOI} \doi{10.1029/1999JA900271} \end{APACrefDOI}
\PrintBackRefs{\CurrentBib}

\bibitem [\protect \citeauthoryear {%
{Fukushima}%
, {Shiokawa}%
, {Otsuka}%
\BCBL {}\ \BBA {} {Ogawa}%
}{%
{Fukushima}%
\ \protect \BOthers {.}}{%
{\protect \APACyear {2012}}%
}]{%
Fukushima2012JGRA..11710324F}
\APACinsertmetastar {%
Fukushima2012JGRA..11710324F}%
\begin{APACrefauthors}%
{Fukushima}, D.%
, {Shiokawa}, K.%
, {Otsuka}, Y.%
\BCBL {}\ \BBA {} {Ogawa}, T.%
\end{APACrefauthors}%
\unskip\
\newblock
\APACrefYearMonthDay{2012}{{\APACmonth{10}}}{}.
\newblock
{\BBOQ}\APACrefatitle {{Observation of equatorial nighttime medium-scale
  traveling ionospheric disturbances in 630-nm airglow images over 7 years}}
  {{Observation of equatorial nighttime medium-scale traveling ionospheric
  disturbances in 630-nm airglow images over 7 years}}.{\BBCQ}
\newblock
\APACjournalVolNumPages{Journal of Geophysical Research (Space
  Physics)}{117}{A10}{A10324}.
\newblock
\begin{APACrefDOI} \doi{10.1029/2012JA017758} \end{APACrefDOI}
\PrintBackRefs{\CurrentBib}

\bibitem [\protect \citeauthoryear {%
Harris%
\ \protect \BOthers {.}}{%
Harris%
\ \protect \BOthers {.}}{%
{\protect \APACyear {2020}}%
}]{%
Numpy2020array}
\APACinsertmetastar {%
Numpy2020array}%
\begin{APACrefauthors}%
Harris, C\BPBI R.%
, Millman, K\BPBI J.%
, van~der Walt, S\BPBI J.%
, Gommers, R.%
, Virtanen, P.%
, Cournapeau, D.%
\BDBL {}Oliphant, T\BPBI E.%
\end{APACrefauthors}%
\unskip\
\newblock
\APACrefYearMonthDay{2020}{{\APACmonth{09}}}{}.
\newblock
{\BBOQ}\APACrefatitle {Array programming with {NumPy}} {Array programming with
  {NumPy}}.{\BBCQ}
\newblock
\APACjournalVolNumPages{Nature}{585}{7825}{357--362}.
\newblock
\begin{APACrefURL} \url{https://doi.org/10.1038/s41586-020-2649-2}
  \end{APACrefURL}
\newblock
\begin{APACrefDOI} \doi{10.1038/s41586-020-2649-2} \end{APACrefDOI}
\PrintBackRefs{\CurrentBib}

\bibitem [\protect \citeauthoryear {%
Helmboldt%
, Haiducek%
\BCBL {}\ \BBA {} Clarke%
}{%
Helmboldt%
\ \protect \BOthers {.}}{%
{\protect \APACyear {2020}}%
}]{%
Helmboldt2020a}
\APACinsertmetastar {%
Helmboldt2020a}%
\begin{APACrefauthors}%
Helmboldt, J\BPBI F.%
, Haiducek, J\BPBI D.%
\BCBL {}\ \BBA {} Clarke, T\BPBI E.%
\end{APACrefauthors}%
\unskip\
\newblock
\APACrefYearMonthDay{2020}{}{}.
\newblock
{\BBOQ}\APACrefatitle {The Properties and Origins of Corotating Plasmaspheric
  Irregularities as Revealed Through a New Tomographic Technique} {The
  properties and origins of corotating plasmaspheric irregularities as revealed
  through a new tomographic technique}.{\BBCQ}
\newblock
\APACjournalVolNumPages{Journal of Geophysical Research: Space
  Physics}{125}{3}{e2019JA027483}.
\newblock
\begin{APACrefURL}
  \url{https://agupubs.onlinelibrary.wiley.com/doi/abs/10.1029/2019JA027483}
  \end{APACrefURL}
\newblock
\APACrefnote{e2019JA027483 10.1029/2019JA027483}
\newblock
\begin{APACrefDOI} \doi{https://doi.org/10.1029/2019JA027483} \end{APACrefDOI}
\PrintBackRefs{\CurrentBib}

\bibitem [\protect \citeauthoryear {%
{Helmboldt}%
\ \BBA {} {Hurley-Walker}%
}{%
{Helmboldt}%
\ \BBA {} {Hurley-Walker}%
}{%
{\protect \APACyear {2020}}%
}]{%
Helm2020RaSc...5507106H}
\APACinsertmetastar {%
Helm2020RaSc...5507106H}%
\begin{APACrefauthors}%
{Helmboldt}, J\BPBI F.%
\BCBT {}\ \BBA {} {Hurley-Walker}, N.%
\end{APACrefauthors}%
\unskip\
\newblock
\APACrefYearMonthDay{2020}{{\APACmonth{10}}}{}.
\newblock
{\BBOQ}\APACrefatitle {{Ionospheric Irregularities Observed During the GLEAM
  Survey}} {{Ionospheric Irregularities Observed During the GLEAM
  Survey}}.{\BBCQ}
\newblock
\APACjournalVolNumPages{Radio Science}{55}{10}{e07106}.
\newblock
\begin{APACrefDOI} \doi{10.1029/2020RS007106} \end{APACrefDOI}
\PrintBackRefs{\CurrentBib}

\bibitem [\protect \citeauthoryear {%
{Helmboldt}%
, {Lane}%
\BCBL {}\ \BBA {} {Cotton}%
}{%
{Helmboldt}%
\ \protect \BOthers {.}}{%
{\protect \APACyear {2012}}%
}]{%
Helmboldt2012RaSc...47.5008H}
\APACinsertmetastar {%
Helmboldt2012RaSc...47.5008H}%
\begin{APACrefauthors}%
{Helmboldt}, J\BPBI F.%
, {Lane}, W\BPBI M.%
\BCBL {}\ \BBA {} {Cotton}, W\BPBI D.%
\end{APACrefauthors}%
\unskip\
\newblock
\APACrefYearMonthDay{2012}{{\APACmonth{10}}}{}.
\newblock
{\BBOQ}\APACrefatitle {{Climatology of midlatitude ionospheric disturbances
  from the Very Large Array Low-frequency Sky Survey}} {{Climatology of
  midlatitude ionospheric disturbances from the Very Large Array Low-frequency
  Sky Survey}}.{\BBCQ}
\newblock
\APACjournalVolNumPages{Radio Science}{47}{5}{RS5008}.
\newblock
\begin{APACrefDOI} \doi{10.1029/2012RS005025} \end{APACrefDOI}
\PrintBackRefs{\CurrentBib}

\bibitem [\protect \citeauthoryear {%
Helmboldt%
, Lazio%
, Intema%
\BCBL {}\ \BBA {} Dymond%
}{%
Helmboldt%
\ \protect \BOthers {.}}{%
{\protect \APACyear {2012}}%
{\protect \APACexlab {{\protect \BCnt {1}}}}}]{%
Helm2012RaSc...47.0K02H_temporal}
\APACinsertmetastar {%
Helm2012RaSc...47.0K02H_temporal}%
\begin{APACrefauthors}%
Helmboldt, J\BPBI F.%
, Lazio, T\BPBI J\BPBI W.%
, Intema, H\BPBI T.%
\BCBL {}\ \BBA {} Dymond, K\BPBI F.%
\end{APACrefauthors}%
\unskip\
\newblock
\APACrefYearMonthDay{2012{\protect \BCnt {1}}}{{\APACmonth{02}}}{}.
\newblock
{\BBOQ}\APACrefatitle {{High-precision measurements of ionospheric TEC
  gradients with the Very Large Array VHF system}} {{High-precision
  measurements of ionospheric TEC gradients with the Very Large Array VHF
  system}}.{\BBCQ}
\newblock
\APACjournalVolNumPages{Radio Science}{47}{}{RS0K02}.
\newblock
\begin{APACrefDOI} \doi{10.1029/2011RS004883} \end{APACrefDOI}
\PrintBackRefs{\CurrentBib}

\bibitem [\protect \citeauthoryear {%
Helmboldt%
, Lazio%
, Intema%
\BCBL {}\ \BBA {} Dymond%
}{%
Helmboldt%
\ \protect \BOthers {.}}{%
{\protect \APACyear {2012}}%
{\protect \APACexlab {{\protect \BCnt {2}}}}}]{%
Helm2012RaSc...47.0L02H_spectral}
\APACinsertmetastar {%
Helm2012RaSc...47.0L02H_spectral}%
\begin{APACrefauthors}%
Helmboldt, J\BPBI F.%
, Lazio, T\BPBI J\BPBI W.%
, Intema, H\BPBI T.%
\BCBL {}\ \BBA {} Dymond, K\BPBI F.%
\end{APACrefauthors}%
\unskip\
\newblock
\APACrefYearMonthDay{2012{\protect \BCnt {2}}}{{\APACmonth{02}}}{}.
\newblock
{\BBOQ}\APACrefatitle {{A new technique for spectral analysis of ionospheric
  TEC fluctuations observed with the Very Large Array VHF system: From QP
  echoes to MSTIDs}} {{A new technique for spectral analysis of ionospheric TEC
  fluctuations observed with the Very Large Array VHF system: From QP echoes to
  MSTIDs}}.{\BBCQ}
\newblock
\APACjournalVolNumPages{Radio Science}{47}{}{RS0L02}.
\newblock
\begin{APACrefDOI} \doi{10.1029/2011RS004787} \end{APACrefDOI}
\PrintBackRefs{\CurrentBib}

\bibitem [\protect \citeauthoryear {%
{Helmboldt}%
\ \BBA {} {Zabotin}%
}{%
{Helmboldt}%
\ \BBA {} {Zabotin}%
}{%
{\protect \APACyear {2022}}%
}]{%
Helmboldt_2022}
\APACinsertmetastar {%
Helmboldt_2022}%
\begin{APACrefauthors}%
{Helmboldt}, J\BPBI F.%
\BCBT {}\ \BBA {} {Zabotin}, N.%
\end{APACrefauthors}%
\unskip\
\newblock
\APACrefYearMonthDay{2022}{{\APACmonth{05}}}{}.
\newblock
{\BBOQ}\APACrefatitle {{An Observed Trend Between Mid-Latitudes Km-Scale
  Irregularities and Medium-Scale Traveling Ionospheric Disturbances}} {{An
  Observed Trend Between Mid-Latitudes Km-Scale Irregularities and Medium-Scale
  Traveling Ionospheric Disturbances}}.{\BBCQ}
\newblock
\APACjournalVolNumPages{Radio Science}{57}{5}{e2021RS007396}.
\newblock
\begin{APACrefDOI} \doi{10.1029/2021RS007396} \end{APACrefDOI}
\PrintBackRefs{\CurrentBib}

\bibitem [\protect \citeauthoryear {%
{Hern{\'a}Ndez-Pajares}%
, {Juan}%
\BCBL {}\ \BBA {} {Sanz}%
}{%
{Hern{\'a}Ndez-Pajares}%
\ \protect \BOthers {.}}{%
{\protect \APACyear {2006}}%
}]{%
MSTID_gps}
\APACinsertmetastar {%
MSTID_gps}%
\begin{APACrefauthors}%
{Hern{\'a}Ndez-Pajares}, M.%
, {Juan}, J\BPBI M.%
\BCBL {}\ \BBA {} {Sanz}, J.%
\end{APACrefauthors}%
\unskip\
\newblock
\APACrefYearMonthDay{2006}{{\APACmonth{07}}}{}.
\newblock
{\BBOQ}\APACrefatitle {{Medium-scale traveling ionospheric disturbances
  affecting GPS measurements: Spatial and temporal analysis}} {{Medium-scale
  traveling ionospheric disturbances affecting GPS measurements: Spatial and
  temporal analysis}}.{\BBCQ}
\newblock
\APACjournalVolNumPages{Journal of Geophysical Research (Space
  Physics)}{111}{A7}{A07S11}.
\newblock
\begin{APACrefDOI} \doi{10.1029/2005JA011474} \end{APACrefDOI}
\PrintBackRefs{\CurrentBib}

\bibitem [\protect \citeauthoryear {%
{Hern{\'a}Ndez-Pajares}%
, {Juan}%
, {Sanz}%
\BCBL {}\ \BBA {} {Arag{\'o}n-{\`A}Ngel}%
}{%
{Hern{\'a}Ndez-Pajares}%
\ \protect \BOthers {.}}{%
{\protect \APACyear {2012}}%
}]{%
Hern2012RaSc...47.0K05H}
\APACinsertmetastar {%
Hern2012RaSc...47.0K05H}%
\begin{APACrefauthors}%
{Hern{\'a}Ndez-Pajares}, M.%
, {Juan}, J\BPBI M.%
, {Sanz}, J.%
\BCBL {}\ \BBA {} {Arag{\'o}n-{\`A}Ngel}, A.%
\end{APACrefauthors}%
\unskip\
\newblock
\APACrefYearMonthDay{2012}{{\APACmonth{08}}}{}.
\newblock
{\BBOQ}\APACrefatitle {{Propagation of medium scale traveling ionospheric
  disturbances at different latitudes and solar cycle conditions}}
  {{Propagation of medium scale traveling ionospheric disturbances at different
  latitudes and solar cycle conditions}}.{\BBCQ}
\newblock
\APACjournalVolNumPages{Radio Science}{47}{4}{RS0K05}.
\newblock
\begin{APACrefDOI} \doi{10.1029/2011RS004951} \end{APACrefDOI}
\PrintBackRefs{\CurrentBib}

\bibitem [\protect \citeauthoryear {%
{Hunter}%
}{%
{Hunter}%
}{%
{\protect \APACyear {2007}}%
}]{%
matplotlib07}
\APACinsertmetastar {%
matplotlib07}%
\begin{APACrefauthors}%
{Hunter}, J\BPBI D.%
\end{APACrefauthors}%
\unskip\
\newblock
\APACrefYearMonthDay{2007}{{\APACmonth{05}}}{}.
\newblock
{\BBOQ}\APACrefatitle {{Matplotlib: A 2D Graphics Environment}} {{Matplotlib: A
  2D Graphics Environment}}.{\BBCQ}
\newblock
\APACjournalVolNumPages{Computing in Science and Engineering}{9}{3}{90-95}.
\newblock
\begin{APACrefDOI} \doi{10.1109/MCSE.2007.55} \end{APACrefDOI}
\PrintBackRefs{\CurrentBib}

\bibitem [\protect \citeauthoryear {%
Ichihara%
, Nishitani%
, Ogawa%
\BCBL {}\ \BBA {} Tsugawa%
}{%
Ichihara%
\ \protect \BOthers {.}}{%
{\protect \APACyear {2013}}%
}]{%
Ichihara201342}
\APACinsertmetastar {%
Ichihara201342}%
\begin{APACrefauthors}%
Ichihara, A.%
, Nishitani, N.%
, Ogawa, T.%
\BCBL {}\ \BBA {} Tsugawa, T.%
\end{APACrefauthors}%
\unskip\
\newblock
\APACrefYearMonthDay{2013}{}{}.
\newblock
{\BBOQ}\APACrefatitle {Northward-propagating nighttime medium-scale traveling
  ionospheric disturbances observed with SuperDARN Hokkaido HF radar and
  GEONET} {Northward-propagating nighttime medium-scale traveling ionospheric
  disturbances observed with superdarn hokkaido hf radar and geonet}{\BBCQ}\
  [Article].
\newblock
\APACjournalVolNumPages{Adv Polar Sci}{24}{1}{42 – 49}.
\newblock
\begin{APACrefURL}
  \url{https://www.scopus.com/inward/record.uri?eid=2-s2.0-85015638139&partnerID=40&md5=d91017f531baac1fbe1697f3e69183a6}
  \end{APACrefURL}
\newblock
\APACrefnote{Cited by: 11}
\PrintBackRefs{\CurrentBib}

\bibitem [\protect \citeauthoryear {%
{Jacobson}%
\ \BBA {} {Erickson}%
}{%
{Jacobson}%
\ \BBA {} {Erickson}%
}{%
{\protect \APACyear {1992}}%
}]{%
Jac1992A&A...257..401J}
\APACinsertmetastar {%
Jac1992A&A...257..401J}%
\begin{APACrefauthors}%
{Jacobson}, A\BPBI R.%
\BCBT {}\ \BBA {} {Erickson}, W\BPBI C.%
\end{APACrefauthors}%
\unskip\
\newblock
\APACrefYearMonthDay{1992}{{\APACmonth{04}}}{}.
\newblock
{\BBOQ}\APACrefatitle {{A method for characterizing transient ionospheric
  disturbances using a large radio telescope array}} {{A method for
  characterizing transient ionospheric disturbances using a large radio
  telescope array}}.{\BBCQ}
\newblock
\APACjournalVolNumPages{Astronomy and Astrophysics}{257}{1}{401-409}.
\PrintBackRefs{\CurrentBib}

\bibitem [\protect \citeauthoryear {%
{Kotake}%
, {Otsuka}%
, {Ogawa}%
, {Tsugawa}%
\BCBL {}\ \BBA {} {Saito}%
}{%
{Kotake}%
\ \protect \BOthers {.}}{%
{\protect \APACyear {2007}}%
}]{%
Kotake2007EP&S...59...95K}
\APACinsertmetastar {%
Kotake2007EP&S...59...95K}%
\begin{APACrefauthors}%
{Kotake}, N.%
, {Otsuka}, Y.%
, {Ogawa}, T.%
, {Tsugawa}, T.%
\BCBL {}\ \BBA {} {Saito}, A.%
\end{APACrefauthors}%
\unskip\
\newblock
\APACrefYearMonthDay{2007}{{\APACmonth{02}}}{}.
\newblock
{\BBOQ}\APACrefatitle {{Statistical study of medium-scale traveling ionospheric
  disturbances observed with the GPS networks in Southern California}}
  {{Statistical study of medium-scale traveling ionospheric disturbances
  observed with the GPS networks in Southern California}}.{\BBCQ}
\newblock
\APACjournalVolNumPages{Earth, Planets and Space}{59}{}{95-102}.
\newblock
\begin{APACrefDOI} \doi{10.1186/BF03352681} \end{APACrefDOI}
\PrintBackRefs{\CurrentBib}

\bibitem [\protect \citeauthoryear {%
{Loi}%
, {Cairns}%
\BCBL {}\ \protect \BOthers {.}}{%
{Loi}%
, {Cairns}%
\BCBL {}\ \protect \BOthers {.}}{%
{\protect \APACyear {2016}}%
}]{%
Loi2016JGRA..121.1569L}
\APACinsertmetastar {%
Loi2016JGRA..121.1569L}%
\begin{APACrefauthors}%
{Loi}, S\BPBI T.%
, {Cairns}, I\BPBI H.%
, {Murphy}, T.%
, {Erickson}, P\BPBI J.%
, {Bell}, M\BPBI E.%
, {Rowlinson}, A.%
\BDBL {}{Kaplan}, D\BPBI L.%
\end{APACrefauthors}%
\unskip\
\newblock
\APACrefYearMonthDay{2016}{{\APACmonth{02}}}{}.
\newblock
{\BBOQ}\APACrefatitle {{Density duct formation in the wake of a travelling
  ionospheric disturbance: Murchison Widefield Array observations}} {{Density
  duct formation in the wake of a travelling ionospheric disturbance: Murchison
  Widefield Array observations}}.{\BBCQ}
\newblock
\APACjournalVolNumPages{Journal of Geophysical Research (Space
  Physics)}{121}{2}{1569-1586}.
\newblock
\begin{APACrefDOI} \doi{10.1002/2015JA022052} \end{APACrefDOI}
\PrintBackRefs{\CurrentBib}

\bibitem [\protect \citeauthoryear {%
{Loi}%
, {Murphy}%
\BCBL {}\ \protect \BOthers {.}}{%
{Loi}%
, {Murphy}%
\BCBL {}\ \protect \BOthers {.}}{%
{\protect \APACyear {2015}}%
}]{%
Loi2015GeoRL..42.3707L}
\APACinsertmetastar {%
Loi2015GeoRL..42.3707L}%
\begin{APACrefauthors}%
{Loi}, S\BPBI T.%
, {Murphy}, T.%
, {Cairns}, I\BPBI H.%
, {Menk}, F\BPBI W.%
, {Waters}, C\BPBI L.%
, {Erickson}, P\BPBI J.%
\BDBL {}{Williams}, C\BPBI L.%
\end{APACrefauthors}%
\unskip\
\newblock
\APACrefYearMonthDay{2015}{{\APACmonth{05}}}{}.
\newblock
{\BBOQ}\APACrefatitle {{Real-time imaging of density ducts between the
  plasmasphere and ionosphere}} {{Real-time imaging of density ducts between
  the plasmasphere and ionosphere}}.{\BBCQ}
\newblock
\APACjournalVolNumPages{Geophysical Research Letters}{42}{10}{3707-3714}.
\newblock
\begin{APACrefDOI} \doi{10.1002/2015GL063699} \end{APACrefDOI}
\PrintBackRefs{\CurrentBib}

\bibitem [\protect \citeauthoryear {%
{Loi}%
, {Murphy}%
\BCBL {}\ \protect \BOthers {.}}{%
{Loi}%
, {Murphy}%
\BCBL {}\ \protect \BOthers {.}}{%
{\protect \APACyear {2016}}%
}]{%
Loi2016RaSc...51..659L}
\APACinsertmetastar {%
Loi2016RaSc...51..659L}%
\begin{APACrefauthors}%
{Loi}, S\BPBI T.%
, {Murphy}, T.%
, {Cairns}, I\BPBI H.%
, {Trott}, C\BPBI M.%
, {Hurley-Walker}, N.%
, {Feng}, L.%
\BDBL {}{Kaplan}, D\BPBI L.%
\end{APACrefauthors}%
\unskip\
\newblock
\APACrefYearMonthDay{2016}{{\APACmonth{06}}}{}.
\newblock
{\BBOQ}\APACrefatitle {{A new angle for probing field-aligned irregularities
  with the Murchison Widefield Array}} {{A new angle for probing field-aligned
  irregularities with the Murchison Widefield Array}}.{\BBCQ}
\newblock
\APACjournalVolNumPages{Radio Science}{51}{6}{659-679}.
\newblock
\begin{APACrefDOI} \doi{10.1002/2015RS005878} \end{APACrefDOI}
\PrintBackRefs{\CurrentBib}

\bibitem [\protect \citeauthoryear {%
{Loi}%
, {Trott}%
\BCBL {}\ \protect \BOthers {.}}{%
{Loi}%
, {Trott}%
\BCBL {}\ \protect \BOthers {.}}{%
{\protect \APACyear {2015}}%
}]{%
Loi_2015}
\APACinsertmetastar {%
Loi_2015}%
\begin{APACrefauthors}%
{Loi}, S\BPBI T.%
, {Trott}, C\BPBI M.%
, {Murphy}, T.%
, {Cairns}, I\BPBI H.%
, {Bell}, M.%
, {Hurley-Walker}, N.%
\BDBL {}{Williams}, C\BPBI L.%
\end{APACrefauthors}%
\unskip\
\newblock
\APACrefYearMonthDay{2015}{{\APACmonth{07}}}{}.
\newblock
{\BBOQ}\APACrefatitle {{Power spectrum analysis of ionospheric fluctuations
  with the Murchison Widefield Array}} {{Power spectrum analysis of ionospheric
  fluctuations with the Murchison Widefield Array}}.{\BBCQ}
\newblock
\APACjournalVolNumPages{Radio Science}{50}{7}{574-597}.
\newblock
\begin{APACrefDOI} \doi{10.1002/2015RS005711} \end{APACrefDOI}
\PrintBackRefs{\CurrentBib}

\bibitem [\protect \citeauthoryear {%
{Mangla}%
, {Chakraborty}%
, {Datta}%
\BCBL {}\ \BBA {} {Paul}%
}{%
{Mangla}%
\ \protect \BOthers {.}}{%
{\protect \APACyear {2023}}%
}]{%
Mangla2022JoAA}
\APACinsertmetastar {%
Mangla2022JoAA}%
\begin{APACrefauthors}%
{Mangla}, S.%
, {Chakraborty}, S.%
, {Datta}, A.%
\BCBL {}\ \BBA {} {Paul}, A.%
\end{APACrefauthors}%
\unskip\
\newblock
\APACrefYearMonthDay{2023}{{\APACmonth{06}}}{}.
\newblock
{\BBOQ}\APACrefatitle {{Exploring Earth's ionosphere and its effect on low
  radio frequency observation with the uGMRT and the SKA}} {{Exploring Earth's
  ionosphere and its effect on low radio frequency observation with the uGMRT
  and the SKA}}.{\BBCQ}
\newblock
\APACjournalVolNumPages{Journal of Astrophysics and Astronomy}{44}{1}{2}.
\newblock
\begin{APACrefDOI} \doi{10.1007/s12036-022-09900-0} \end{APACrefDOI}
\PrintBackRefs{\CurrentBib}

\bibitem [\protect \citeauthoryear {%
{Mangla}%
\ \BBA {} {Datta}%
}{%
{Mangla}%
\ \BBA {} {Datta}%
}{%
{\protect \APACyear {2022}}%
}]{%
Mangla2022MNRAS.513..964M}
\APACinsertmetastar {%
Mangla2022MNRAS.513..964M}%
\begin{APACrefauthors}%
{Mangla}, S.%
\BCBT {}\ \BBA {} {Datta}, A.%
\end{APACrefauthors}%
\unskip\
\newblock
\APACrefYearMonthDay{2022}{{\APACmonth{06}}}{}.
\newblock
{\BBOQ}\APACrefatitle {{Study of the equatorial ionosphere using the Giant
  Metrewave Radio Telescope (GMRT) at sub-GHz frequencies}} {{Study of the
  equatorial ionosphere using the Giant Metrewave Radio Telescope (GMRT) at
  sub-GHz frequencies}}.{\BBCQ}
\newblock
\APACjournalVolNumPages{Monthly Notices of the Royal Astronomical
  Society}{513}{1}{964-975}.
\newblock
\begin{APACrefDOI} \doi{10.1093/mnras/stac942} \end{APACrefDOI}
\PrintBackRefs{\CurrentBib}

\bibitem [\protect \citeauthoryear {%
Mevius%
\ \protect \BOthers {.}}{%
Mevius%
\ \protect \BOthers {.}}{%
{\protect \APACyear {2016}}%
}]{%
mev16}
\APACinsertmetastar {%
mev16}%
\begin{APACrefauthors}%
Mevius, M.%
, van~der Tol, S.%
, Pandey, V\BPBI N.%
, Vedantham, H\BPBI K.%
, Brentjens, M\BPBI A.%
, de Bruyn, A\BPBI G.%
\BDBL {}Zaroubi, S.%
\end{APACrefauthors}%
\unskip\
\newblock
\APACrefYearMonthDay{2016}{}{}.
\newblock
{\BBOQ}\APACrefatitle {Probing ionospheric structures using the LOFAR radio
  telescope} {Probing ionospheric structures using the lofar radio
  telescope}.{\BBCQ}
\newblock
\APACjournalVolNumPages{Radio Science}{51}{7}{927-941}.
\newblock
\begin{APACrefURL}
  \url{https://agupubs.onlinelibrary.wiley.com/doi/abs/10.1002/2016RS006028}
  \end{APACrefURL}
\newblock
\begin{APACrefDOI} \doi{10.1002/2016RS006028} \end{APACrefDOI}
\PrintBackRefs{\CurrentBib}

\bibitem [\protect \citeauthoryear {%
Mrak%
, Semeter%
, Nishimura%
\BCBL {}\ \BBA {} Coster%
}{%
Mrak%
\ \protect \BOthers {.}}{%
{\protect \APACyear {2021}}%
}]{%
Mrak2021}
\APACinsertmetastar {%
Mrak2021}%
\begin{APACrefauthors}%
Mrak, S.%
, Semeter, J.%
, Nishimura, Y.%
\BCBL {}\ \BBA {} Coster, A\BPBI J.%
\end{APACrefauthors}%
\unskip\
\newblock
\APACrefYearMonthDay{2021}{}{}.
\newblock
{\BBOQ}\APACrefatitle {Extreme Low-Latitude Total Electron Content Enhancement
  and Global Positioning System Scintillation at Dawn} {Extreme low-latitude
  total electron content enhancement and global positioning system
  scintillation at dawn}.{\BBCQ}
\newblock
\APACjournalVolNumPages{Space Weather}{19}{9}{e2021SW002740}.
\newblock
\begin{APACrefURL}
  \url{https://agupubs.onlinelibrary.wiley.com/doi/abs/10.1029/2021SW002740}
  \end{APACrefURL}
\newblock
\APACrefnote{e2021SW002740 2021SW002740}
\newblock
\begin{APACrefDOI} \doi{https://doi.org/10.1029/2021SW002740} \end{APACrefDOI}
\PrintBackRefs{\CurrentBib}

\bibitem [\protect \citeauthoryear {%
{Mukherjee}%
, {R}%
, {Parihar}%
, {Ghodpage}%
\BCBL {}\ \BBA {} {Patil}%
}{%
{Mukherjee}%
\ \protect \BOthers {.}}{%
{\protect \APACyear {2010}}%
}]{%
Mukherjee_2010}
\APACinsertmetastar {%
Mukherjee_2010}%
\begin{APACrefauthors}%
{Mukherjee}, G\BPBI K.%
, {R}, P\BPBI S.%
, {Parihar}, N.%
, {Ghodpage}, R.%
\BCBL {}\ \BBA {} {Patil}, P\BPBI T.%
\end{APACrefauthors}%
\unskip\
\newblock
\APACrefYearMonthDay{2010}{{\APACmonth{03}}}{}.
\newblock
{\BBOQ}\APACrefatitle {{Studies of the wind filtering effect of gravity waves
  observed at Allahabad (25.45{\textdegree}N, 81.85{\textdegree}E) in India}}
  {{Studies of the wind filtering effect of gravity waves observed at Allahabad
  (25.45{\textdegree}N, 81.85{\textdegree}E) in India}}.{\BBCQ}
\newblock
\APACjournalVolNumPages{Earth, Planets and Space}{62}{3}{309-318}.
\newblock
\begin{APACrefDOI} \doi{10.5047/eps.2009.11.008} \end{APACrefDOI}
\PrintBackRefs{\CurrentBib}

\bibitem [\protect \citeauthoryear {%
{Narayanan}%
, {Shiokawa}%
, {Otsuka}%
\BCBL {}\ \BBA {} {Saito}%
}{%
{Narayanan}%
\ \protect \BOthers {.}}{%
{\protect \APACyear {2014}}%
}]{%
Narayanan2014JGRA..119.9268L}
\APACinsertmetastar {%
Narayanan2014JGRA..119.9268L}%
\begin{APACrefauthors}%
{Narayanan}, V\BPBI L\BPBI L.%
, {Shiokawa}, K.%
, {Otsuka}, Y.%
\BCBL {}\ \BBA {} {Saito}, S.%
\end{APACrefauthors}%
\unskip\
\newblock
\APACrefYearMonthDay{2014}{{\APACmonth{11}}}{}.
\newblock
{\BBOQ}\APACrefatitle {{Airglow observations of nighttime medium-scale
  traveling ionospheric disturbances from Yonaguni: Statistical characteristics
  and low-latitude limit}} {{Airglow observations of nighttime medium-scale
  traveling ionospheric disturbances from Yonaguni: Statistical characteristics
  and low-latitude limit}}.{\BBCQ}
\newblock
\APACjournalVolNumPages{Journal of Geophysical Research (Space
  Physics)}{119}{11}{9268-9282}.
\newblock
\begin{APACrefDOI} \doi{10.1002/2014JA020368} \end{APACrefDOI}
\PrintBackRefs{\CurrentBib}

\bibitem [\protect \citeauthoryear {%
{Otsuka}%
\ \protect \BOthers {.}}{%
{Otsuka}%
\ \protect \BOthers {.}}{%
{\protect \APACyear {2013}}%
}]{%
Otsuka2013AnGeo..31..163O}
\APACinsertmetastar {%
Otsuka2013AnGeo..31..163O}%
\begin{APACrefauthors}%
{Otsuka}, Y.%
, {Suzuki}, K.%
, {Nakagawa}, S.%
, {Nishioka}, M.%
, {Shiokawa}, K.%
\BCBL {}\ \BBA {} {Tsugawa}, T.%
\end{APACrefauthors}%
\unskip\
\newblock
\APACrefYearMonthDay{2013}{{\APACmonth{02}}}{}.
\newblock
{\BBOQ}\APACrefatitle {{GPS observations of medium-scale traveling ionospheric
  disturbances over Europe}} {{GPS observations of medium-scale traveling
  ionospheric disturbances over Europe}}.{\BBCQ}
\newblock
\APACjournalVolNumPages{Annales Geophysicae}{31}{2}{163-172}.
\newblock
\begin{APACrefDOI} \doi{10.5194/angeo-31-163-2013} \end{APACrefDOI}
\PrintBackRefs{\CurrentBib}

\bibitem [\protect \citeauthoryear {%
{Pacheco}%
, {Heelis}%
\BCBL {}\ \BBA {} {Su}%
}{%
{Pacheco}%
\ \protect \BOthers {.}}{%
{\protect \APACyear {2011}}%
}]{%
Pacheco2011JGRA..11611329P}
\APACinsertmetastar {%
Pacheco2011JGRA..11611329P}%
\begin{APACrefauthors}%
{Pacheco}, E\BPBI E.%
, {Heelis}, R\BPBI A.%
\BCBL {}\ \BBA {} {Su}, S\BPBI Y.%
\end{APACrefauthors}%
\unskip\
\newblock
\APACrefYearMonthDay{2011}{{\APACmonth{11}}}{}.
\newblock
{\BBOQ}\APACrefatitle {{Superrotation of the ionosphere and quiet time zonal
  ion drifts at low and middle latitudes observed by Republic of China
  Satellite-1 (ROCSAT-1)}} {{Superrotation of the ionosphere and quiet time
  zonal ion drifts at low and middle latitudes observed by Republic of China
  Satellite-1 (ROCSAT-1)}}.{\BBCQ}
\newblock
\APACjournalVolNumPages{Journal of Geophysical Research (Space
  Physics)}{116}{A11}{A11329}.
\newblock
\begin{APACrefDOI} \doi{10.1029/2011JA016786} \end{APACrefDOI}
\PrintBackRefs{\CurrentBib}

\bibitem [\protect \citeauthoryear {%
{Shiokawa}%
, {Ihara}%
, {Otsuka}%
\BCBL {}\ \BBA {} {Ogawa}%
}{%
{Shiokawa}%
\ \protect \BOthers {.}}{%
{\protect \APACyear {2003}}%
}]{%
Shiokawa2003JGRA..108.1052S}
\APACinsertmetastar {%
Shiokawa2003JGRA..108.1052S}%
\begin{APACrefauthors}%
{Shiokawa}, K.%
, {Ihara}, C.%
, {Otsuka}, Y.%
\BCBL {}\ \BBA {} {Ogawa}, T.%
\end{APACrefauthors}%
\unskip\
\newblock
\APACrefYearMonthDay{2003}{{\APACmonth{01}}}{}.
\newblock
{\BBOQ}\APACrefatitle {{Statistical study of nighttime medium-scale traveling
  ionospheric disturbances using midlatitude airglow images}} {{Statistical
  study of nighttime medium-scale traveling ionospheric disturbances using
  midlatitude airglow images}}.{\BBCQ}
\newblock
\APACjournalVolNumPages{Journal of Geophysical Research (Space
  Physics)}{108}{A1}{1052}.
\newblock
\begin{APACrefDOI} \doi{10.1029/2002JA009491} \end{APACrefDOI}
\PrintBackRefs{\CurrentBib}

\bibitem [\protect \citeauthoryear {%
Shiokawa%
, Otsuka%
\BCBL {}\ \BBA {} Ogawa%
}{%
Shiokawa%
\ \protect \BOthers {.}}{%
{\protect \APACyear {2006}}%
}]{%
Shiokawa_2005JA011406}
\APACinsertmetastar {%
Shiokawa_2005JA011406}%
\begin{APACrefauthors}%
Shiokawa, K.%
, Otsuka, Y.%
\BCBL {}\ \BBA {} Ogawa, T.%
\end{APACrefauthors}%
\unskip\
\newblock
\APACrefYearMonthDay{2006}{}{}.
\newblock
{\BBOQ}\APACrefatitle {Quasiperiodic southward moving waves in 630-nm airglow
  images in the equatorial thermosphere} {Quasiperiodic southward moving waves
  in 630-nm airglow images in the equatorial thermosphere}.{\BBCQ}
\newblock
\APACjournalVolNumPages{Journal of Geophysical Research: Space
  Physics}{111}{A6}{}.
\newblock
\begin{APACrefURL}
  \url{https://agupubs.onlinelibrary.wiley.com/doi/abs/10.1029/2005JA011406}
  \end{APACrefURL}
\newblock
\begin{APACrefDOI} \doi{https://doi.org/10.1029/2005JA011406} \end{APACrefDOI}
\PrintBackRefs{\CurrentBib}

\bibitem [\protect \citeauthoryear {%
{Sivakandan}%
\ \protect \BOthers {.}}{%
{Sivakandan}%
\ \protect \BOthers {.}}{%
{\protect \APACyear {2019}}%
}]{%
Sivakandan2019JGRA..124..749S}
\APACinsertmetastar {%
Sivakandan2019JGRA..124..749S}%
\begin{APACrefauthors}%
{Sivakandan}, M.%
, {Chakrabarty}, D.%
, {Ramkumar}, T\BPBI K.%
, {Guharay}, A.%
, {Taori}, A.%
\BCBL {}\ \BBA {} {Parihar}, N.%
\end{APACrefauthors}%
\unskip\
\newblock
\APACrefYearMonthDay{2019}{{\APACmonth{01}}}{}.
\newblock
{\BBOQ}\APACrefatitle {{Evidence for Deep Ingression of the Midlatitude MSTID
  Into As Low as 3.5{\textdegree} Magnetic Latitude}} {{Evidence for Deep
  Ingression of the Midlatitude MSTID Into As Low as 3.5{\textdegree} Magnetic
  Latitude}}.{\BBCQ}
\newblock
\APACjournalVolNumPages{Journal of Geophysical Research (Space
  Physics)}{124}{1}{749-764}.
\newblock
\begin{APACrefDOI} \doi{10.1029/2018JA026103} \end{APACrefDOI}
\PrintBackRefs{\CurrentBib}

\bibitem [\protect \citeauthoryear {%
{Sultan}%
}{%
{Sultan}%
}{%
{\protect \APACyear {1996}}%
}]{%
Sultan1996JGR...10126875S}
\APACinsertmetastar {%
Sultan1996JGR...10126875S}%
\begin{APACrefauthors}%
{Sultan}, P\BPBI J.%
\end{APACrefauthors}%
\unskip\
\newblock
\APACrefYearMonthDay{1996}{{\APACmonth{12}}}{}.
\newblock
{\BBOQ}\APACrefatitle {{Linear theory and modeling of the Rayleigh-Taylor
  instability leading to the occurrence of equatorial spread F}} {{Linear
  theory and modeling of the Rayleigh-Taylor instability leading to the
  occurrence of equatorial spread F}}.{\BBCQ}
\newblock
\APACjournalVolNumPages{Journal of Geophysical Research: Space
  Physics}{101}{A12}{26875-26892}.
\newblock
\begin{APACrefDOI} \doi{10.1029/96JA00682} \end{APACrefDOI}
\PrintBackRefs{\CurrentBib}

\bibitem [\protect \citeauthoryear {%
{Thompson}%
}{%
{Thompson}%
}{%
{\protect \APACyear {1999}}%
}]{%
thompson_book}
\APACinsertmetastar {%
thompson_book}%
\begin{APACrefauthors}%
{Thompson}, A\BPBI R.%
\end{APACrefauthors}%
\unskip\
\newblock
\APACrefYearMonthDay{1999}{}{}.
\newblock
{\BBOQ}\APACrefatitle {{Fundamentals of Radio Interferometry}} {{Fundamentals
  of Radio Interferometry}}.{\BBCQ}
\newblock
\BIn{} G\BPBI B.~{Taylor}, C\BPBI L.~{Carilli}\BCBL {}\ \BBA {} R\BPBI
  A.~{Perley}\ (\BEDS), \APACrefbtitle {Synthesis Imaging in Radio Astronomy
  II} {Synthesis imaging in radio astronomy ii}\ (\BVOL~180, \BPG~11).
\PrintBackRefs{\CurrentBib}

\bibitem [\protect \citeauthoryear {%
{Tsugawa}%
, {Otsuka}%
, {Coster}%
\BCBL {}\ \BBA {} {Saito}%
}{%
{Tsugawa}%
\ \protect \BOthers {.}}{%
{\protect \APACyear {2007}}%
}]{%
Tsugawa2007GeoRL..3422101T}
\APACinsertmetastar {%
Tsugawa2007GeoRL..3422101T}%
\begin{APACrefauthors}%
{Tsugawa}, T.%
, {Otsuka}, Y.%
, {Coster}, A\BPBI J.%
\BCBL {}\ \BBA {} {Saito}, A.%
\end{APACrefauthors}%
\unskip\
\newblock
\APACrefYearMonthDay{2007}{{\APACmonth{11}}}{}.
\newblock
{\BBOQ}\APACrefatitle {{Medium-scale traveling ionospheric disturbances
  detected with dense and wide TEC maps over North America}} {{Medium-scale
  traveling ionospheric disturbances detected with dense and wide TEC maps over
  North America}}.{\BBCQ}
\newblock
\APACjournalVolNumPages{Geophysical Research Letters}{34}{22}{L22101}.
\newblock
\begin{APACrefDOI} \doi{10.1029/2007GL031663} \end{APACrefDOI}
\PrintBackRefs{\CurrentBib}

\bibitem [\protect \citeauthoryear {%
Tsunoda%
\ \BBA {} White%
}{%
Tsunoda%
\ \BBA {} White%
}{%
{\protect \APACyear {1981}}%
}]{%
TsunodaJA086iA05p03610}
\APACinsertmetastar {%
TsunodaJA086iA05p03610}%
\begin{APACrefauthors}%
Tsunoda, R\BPBI T.%
\BCBT {}\ \BBA {} White, B\BPBI R.%
\end{APACrefauthors}%
\unskip\
\newblock
\APACrefYearMonthDay{1981}{}{}.
\newblock
{\BBOQ}\APACrefatitle {On the generation and growth of equatorial backscatter
  plumes 1. Wave structure in the bottomside F layer} {On the generation and
  growth of equatorial backscatter plumes 1. wave structure in the bottomside f
  layer}.{\BBCQ}
\newblock
\APACjournalVolNumPages{Journal of Geophysical Research: Space
  Physics}{86}{A5}{3610-3616}.
\newblock
\begin{APACrefURL}
  \url{https://agupubs.onlinelibrary.wiley.com/doi/abs/10.1029/JA086iA05p03610}
  \end{APACrefURL}
\newblock
\begin{APACrefDOI} \doi{https://doi.org/10.1029/JA086iA05p03610}
  \end{APACrefDOI}
\PrintBackRefs{\CurrentBib}

\bibitem [\protect \citeauthoryear {%
Virtanen%
\ \protect \BOthers {.}}{%
Virtanen%
\ \protect \BOthers {.}}{%
{\protect \APACyear {2020}}%
}]{%
SciPy-NMeth2020}
\APACinsertmetastar {%
SciPy-NMeth2020}%
\begin{APACrefauthors}%
Virtanen, P.%
, Gommers, R.%
, Oliphant, T\BPBI E.%
, Haberland, M.%
, Reddy, T.%
, Cournapeau, D.%
\BDBL {}{SciPy 1.0 Contributors}%
\end{APACrefauthors}%
\unskip\
\newblock
\APACrefYearMonthDay{2020}{}{}.
\newblock
{\BBOQ}\APACrefatitle {{{SciPy} 1.0: Fundamental Algorithms for Scientific
  Computing in Python}} {{{SciPy} 1.0: Fundamental Algorithms for Scientific
  Computing in Python}}.{\BBCQ}
\newblock
\APACjournalVolNumPages{Nature Methods}{17}{}{261--272}.
\newblock
\begin{APACrefDOI} \doi{10.1038/s41592-019-0686-2} \end{APACrefDOI}
\PrintBackRefs{\CurrentBib}

\bibitem [\protect \citeauthoryear {%
Waskom%
}{%
Waskom%
}{%
{\protect \APACyear {2021}}%
}]{%
seaborn2021}
\APACinsertmetastar {%
seaborn2021}%
\begin{APACrefauthors}%
Waskom, M\BPBI L.%
\end{APACrefauthors}%
\unskip\
\newblock
\APACrefYearMonthDay{2021}{}{}.
\newblock
{\BBOQ}\APACrefatitle {seaborn: statistical data visualization} {seaborn:
  statistical data visualization}.{\BBCQ}
\newblock
\APACjournalVolNumPages{Journal of Open Source Software}{6}{60}{3021}.
\newblock
\begin{APACrefURL} \url{https://doi.org/10.21105/joss.03021} \end{APACrefURL}
\newblock
\begin{APACrefDOI} \doi{10.21105/joss.03021} \end{APACrefDOI}
\PrintBackRefs{\CurrentBib}

\bibitem [\protect \citeauthoryear {%
Wayth%
\ \protect \BOthers {.}}{%
Wayth%
\ \protect \BOthers {.}}{%
{\protect \APACyear {2015}}%
}]{%
Wayth2015PASA...32...25W}
\APACinsertmetastar {%
Wayth2015PASA...32...25W}%
\begin{APACrefauthors}%
Wayth, R\BPBI B.%
, Lenc, E.%
, Bell, M\BPBI E.%
, Callingham, J\BPBI R.%
, Dwarakanath, K\BPBI S.%
, Franzen, T\BPBI M\BPBI O.%
\BDBL {}et al.%
\end{APACrefauthors}%
\unskip\
\newblock
\APACrefYearMonthDay{2015}{}{}.
\newblock
{\BBOQ}\APACrefatitle {GLEAM: The GaLactic and Extragalactic All-Sky MWA
  Survey} {Gleam: The galactic and extragalactic all-sky mwa survey}.{\BBCQ}
\newblock
\APACjournalVolNumPages{Publications of the Astronomical Society of
  Australia}{32}{}{e025}.
\newblock
\begin{APACrefDOI} \doi{10.1017/pasa.2015.26} \end{APACrefDOI}
\PrintBackRefs{\CurrentBib}

\end{thebibliography}

\appendix
\section{Higher order effects}
\label{appendix_a}
By comparing equations \ref{eq:FT_theory} and \ref{eq:FT_observe}, we also get the estimators for $k_x$ and $k_y$ from the higher order terms which are as follows
\begin{eqnarray}
    \label{eq:higher_wavenumbers}
    k_x^2(\omega) \simeq - \left(  \frac{6P_5}{P_0}(\omega) \right)  \simeq - \left( \frac{2P_7}{P_1}(\omega) \right) \nonumber \\
    k_y^2(\omega) \simeq - \left(  \frac{6P_6}{P_1}(\omega) \right)  \simeq - \left( \frac{2P_8}{P_0}(\omega) \right) \nonumber \\
    k_x (\omega) k_y(\omega) \simeq -\left( \frac{2P_7}{P_0}(\omega) \right) \simeq -\left( \frac{2P_8}{P_1}(\omega) \right)
\end{eqnarray}
Assuming $k_p^2$ = $A\,e^{i\phi}$, gives

\begin{eqnarray}
\label{eq:example_wavenumber}
    k_p = 
\begin{cases}
    \sqrt{A}\,e^{i\left( \frac{\phi}{2} +n\pi \right)}              & \text{for } n \text{ is } even \\
    \sqrt{A}\,e^{i\left( \frac{\phi}{2} +\frac{n}{2}\pi \right)}    & \text{for } n \text{ is } odd
\end{cases}
\end{eqnarray}
For each $k_x$ and $k_y$, there are two values, one positive and one negative, corresponding to two directions of each wave vector. This creates a problem when calculating the actual propagation direction from equation \ref{eq:azimuth}. To avoid considering higher order effects, these equations are not used to estimate the wave vectors.

\end{document}